\documentclass[10pt,a4paper,english,prd,nofootinbib,notitlepage,superscriptaddress,showkeys]{revtex4-1}
\usepackage{color}
\usepackage{babel}
\usepackage{amsmath}
\usepackage{amssymb}
\usepackage{esint}
\usepackage{hyperref}
\usepackage{breakurl}
\usepackage{cleveref}
\usepackage{ulem}
\usepackage{bbm}

\makeatletter

\providecommand{\tabularnewline}{\\}

\@ifundefined{textcolor}{}
{%
 \definecolor{BLACK}{gray}{0}
 \definecolor{WHITE}{gray}{1}
 \definecolor{RED}{rgb}{1,0,0}
 \definecolor{GREEN}{rgb}{0,1,0}
 \definecolor{BLUE}{rgb}{0,0,1}
 \definecolor{CYAN}{cmyk}{1,0,0,0}
 \definecolor{MAGENTA}{cmyk}{0,1,0,0}
 \definecolor{YELLOW}{cmyk}{0,0,1,0}
}

\makeatother

\begin{document}

\title{Consistent metric combinations in cosmology of massive bigravity}

\author{Henrik Nersisyan}
\email{h.nersisyan@thphys.uni-heidelberg.de}
\affiliation{Institut F\"ur Theoretische Physik, Ruprecht-Karls-Universit\"at
Heidelberg, Philosophenweg 16, 69120 Heidelberg, Germany}
\author{Yashar Akrami}
\email{y.akrami@thphys.uni-heidelberg.de}
\affiliation{Institut F\"ur Theoretische Physik, Ruprecht-Karls-Universit\"at
Heidelberg, Philosophenweg 16, 69120 Heidelberg, Germany}
\author{Luca Amendola}
\email{l.amendola@thphys.uni-heidelberg.de}
\affiliation{Institut F\"ur Theoretische Physik, Ruprecht-Karls-Universit\"at
Heidelberg, Philosophenweg 16, 69120 Heidelberg, Germany}

\begin{abstract}
Massive bigravity models are interesting alternatives to standard
cosmology. In most cases, however, these models have been studied for
a simplified scenario in which both metrics take homogeneous and isotropic
forms [Friedmann-Lema\^{i}tre-Robertson-Walker (FLRW)] with the same spatial curvatures. The interest to consider
more general geometries arises, in particular, in view of the difficulty so far
encountered in building stable cosmological solutions with homogeneous
and isotropic metrics. Here we consider a number of cases in which
the two metrics take more general forms, namely FLRW with different spatial
curvatures---Lema\^{i}tre, Lema\^{i}tre-Tolman-Bondi (LTB), and Bianchi
I---as well as cases where only one metric is linearly perturbed. We discuss possible consistent combinations and find that only some special cases of FLRW--Lema\^{i}tre, LTB--LTB, and
FLRW--Bianchi I combinations give consistent, nontrivial solutions. 
\end{abstract}

\keywords{modified gravity, massive gravity, bimetric gravity, background cosmology}

\maketitle

\section{Introduction}
\label{sec:intro}

The standard $\Lambda$CDM model of cosmology is based on four main
assumptions: general relativity (GR) is the correct description of
gravitational interactions at energies below the Planck scale, the
Universe is homogenous and isotropic on large scales (the cosmological
principle), the energy content of the Universe is mainly in the form
of cold dark matter (CDM) and a nondynamical cosmological constant
$\Lambda$, and all the structure that we see around us originated
from nearly Gaussian, adiabatic, and scale-independent quantum fluctuations
at early times. All these assumptions have been tested with high precision
using various cosmological data and seem to be in excellent agreement
with all existing observations. There are, however, various theoretical
reasons why one may want to go beyond this standard framework. In
particular, the assumption that the late-time acceleration of the
Universe is due to a cosmological constant term has been strongly
questioned from the theoretical point of view, as its small but nonzero
value preferred by observations cannot be explained by fundamental
physics~\cite{Martin:2012bt}. It is, therefore, important and quite
natural to ask whether the cosmic acceleration can be explained by
a different mechanism than a pure cosmological constant. One particular
possibility, which has attracted remarkable attention over the last
decade, is that a modification of GR on very large scales might be
responsible for the acceleration (see Refs.~\cite{2010deto.book.....A,Clifton:2011jh}
for comprehensive reviews). One of the interesting such infrared modifications
is to assume that gravitons are not massless as opposed to what GR
tells us. A nonzero but sufficiently small graviton mass modifies
properties of the gravitational interactions on very large scales
while leaving them indistinguishable from the predictions of standard
gravity on small scales where GR is believed to be at work.

GR is a consistent and nonlinear theory of massless gravity and, therefore,
has given the possibility of constructing various cosmological models.
In order to test the implications of massive gravity for cosmology,
one similarly needs a nonlinear and consistent theory for massive
gravitons. Such a theory was, however, not available for more than 70
years after the construction of a linear theory of massive gravity
by Fierz and Pauli in 1939~\cite{Fierz:1939ix}. This was mainly
because any attempts at constructing a nonlinear completion of the
Fierz and Pauli theory would face a serious obstacle; the theory would
suffer from the existence of the so-called Boulware-Deser (BD) ghost
degrees of freedom~\cite{Boulware:1973my}, a property which would
be fatal to the theory. It was only a few years ago that a ghost-free
and fully nonlinear formulation of massive gravity, and its bimetric
extension, was constructed~\cite{deRham:2010ik,deRham:2010kj,deRham:2011rn,deRham:2011qq,Hassan:2011vm,Hassan:2011hr,Hassan:2011tf,Hassan:2011zd,Hassan:2011ea}
(see Ref.~\cite{deRham:2014zqa} for a recent review). The key step
for this success was to extend the gravitational sector by at least
one new spin-2 tensor field with metric-like properties. In addition,
in order to avoid the BD ghost, the physical metric of the theory
has to interact with the new tensor field in a very specific way.
In the simplest version of the theory, referred to as the de Rham-Gabadadze-Tolley
(dRGT) theory of massive gravity, only the physical metric, the one
which interacts with the matter sector in the standard way, is dynamical,
i.e. has an Einstein-Hilbert term in the action, while the second
metric, often called ``reference'' metric, does not have dynamics.
In this case gravitons posses five degrees of freedom. In the bimetric
version of the theory, referred to as the Hassan-Rosen theory of bigravity,
the reference metric is also given dynamics and, therefore, gravitons
posses seven degrees of freedom, corresponding to one massless and
one massive graviton.

The dRGT theory of massive gravity has been shown to suffer from a
no-go theorem forbidding flat and closed Friedmann-Lema\^{i}tre-Robertson-Walker
(FLRW) cosmological solutions on a flat reference metric~\cite{D'Amico:2011jj}.
In addition, dRGT with an open FLRW metric or a nonflat reference
metric suffers from the so-called Higuchi instability~\cite{Higuchi:1986py}
or other types of instabilities~\cite{Gumrukcuoglu:2011ew,Gumrukcuoglu:2011zh,Vakili:2012tm,DeFelice:2012mx,Fasiello:2012rw,DeFelice:2013awa}.
One obvious way to avoid the no-go theorem is to give up on exact
FLRW solutions, i.e. to consider inhomogeneous and/or anisotropic
solutions for the metrics. We, however, know that the observable Universe
on large scales is very close to being homogenous and isotropic and,
therefore, non-FLRW solutions, if allowed, must not deviate significantly
from the FLRW case on observable scales, and must respect the observational
bounds on inhomogeneity and anisotropy. Non-FLRW effects should be
either of very low amplitudes or of scales much larger than our horizon
so that they cannot be observed. Interestingly, solutions which satisfy
these conditions have been shown to exist in dRGT~\cite{D'Amico:2011jj}.
Other scenarios with non-FLRW solutions in dRGT, either for the physical
metric or for the reference metric, can be found in Refs.~\cite{Gratia:2012wt,Gumrukcuoglu:2012aa,DeFelice:2012mx,Volkov:2012cf,Volkov:2012zb,DeFelice:2013awa,DeFelice:2013bxa,Do:2013tea,Kao:2014efa}
(see also Ref.~\cite{deRham:2014zqa} for a thorough review of the
inhomogeneous and anisotropic solutions in massive gravity). The other
possibility to avoid the no-go theorem and instability issues in dRGT
is to extend the theory. An example for such extensions without adding
new degrees of freedom is the recently proposed generalized massive
gravity theory~\cite{deRham:2014gla}. Another workaround is through
theories with extra degrees of freedom or violation of certain symmetries; these include for example
quasidilaton \cite{D'Amico:2012zv}, varying-mass \cite{D'Amico:2011jj,Huang:2012pe},
nonlocal \cite{Jaccard:2013gla,Foffa:2013vma,Dirian:2014ara,Conroy:2014eja}, and
Lorentz-violating \cite{Comelli:2013paa,Comelli:2013tja} massive
gravity. Recently, another solution to the no-go theorem has been
proposed in Refs.~\cite{deRham:2014naa,deRham:2014fha,Gumrukcuoglu:2014xba}.
It has been suggested that the no-go theorem can be overcome if at
least some matter couples to a hybrid metric, composed of both the
physical and reference metrics and constructed in a specific way to
keep the theory free of the BD ghost up to a cut-off energy scale
which is believed to be above the strong coupling scale of the theory.
This makes phenomenological studies of the theory possible below the
cut-off scale (see, however, Ref.~\cite{Solomon:2014iwa} for various
complications that the cosmology of this theory would need to tackle).
The revival of the BD ghost is a general feature of the scenarios
where both metrics couple simultaneously to matter~\cite{Yamashita:2014fga,deRham:2014naa,Noller:2014sta,Hassan:2014gta,deRham:2014fha,Soloviev:2014eea,Heisenberg:2014rka};
these include the simplest case where the two metrics couple to matter
minimally~\cite{Akrami:2013ffa,Akrami:2014lja,Khosravi:2011zi,Tamanini:2013xia}.

The Hassan-Rosen theory of massive bigravity on the other hand is
immune from the no-go theorem and admits usual flat FLRW solutions.
The cosmology of bigravity has been extensively studied in the literature
at both the background and perturbative levels (see, e.g., Refs.~\cite{Volkov:2011an,vonStrauss:2011mq,Comelli:2011zm,Comelli:2012db,Khosravi:2012rk,Berg:2012kn,Akrami:2012vf,Akrami:2013pna,Enander:2013kza,Fasiello:2013woa,Konnig:2013gxa,Konnig:2014dna,Comelli:2014bqa,DeFelice:2014nja,Solomon:2014dua,Konnig:2014xva,Lagos:2014lca,Cusin:2014psa,Enander:2015vja})
and in terms of the lensing and dynamical properties of local sources~\cite{Enander:2013kza}
(see also Refs.~\cite{Enander:2014xga,Schmidt-May:2014xla,Comelli:2015pua,Gumrukcuoglu:2015nua}
for cosmological studies of bigravity where matter couples to both
metrics through a composite metric). It has particularly been shown
that the theory can explain the late-time acceleration of the Universe
in the absence of an explicit cosmological constant and can, therefore,
serve as a viable alternative to $\Lambda$CDM~\cite{Akrami:2012vf,Konnig:2013gxa}
at the background level. However, when it comes to perturbations,
the Hassan-Rosen theory of massive bigravity seems to be suffering
from various instabilities. Bigravity models can generally be classified
into two categories, finite and infinite branches. This classification
is based on the fact that the ratio of the two scale factors for the
reference and physical metrics in FLRW solutions is either increasing
(finite branch) or decreasing (infinite branch) with time, depending
on which combinations of the parameters of the theory are nonvanishing.
It has been shown that scalar perturbations for all finite-branch
models, including a simple single-parameter model called minimal bigravity
model (MBM)~\cite{Konnig:2014dna}, are unstable at early times on
small scales~\cite{Comelli:2012db,Konnig:2014dna,Konnig:2014xva,DeFelice:2014nja,Lagos:2014lca},
although the models are viable at the background level. The instabilities
do not necessarily rule these models out but make their comparison
to observations difficult as one can no longer employ linear perturbation
theory to study their implications for the formation of structure.
Linear scalar perturbations for the infinite-branch bigravity (IBB),
as identified in Ref.~\cite{Konnig:2014xva}, are on the other hand
stable at all times, making the study of structure formation possible
for the model. This was done in Refs.~\cite{Solomon:2014dua,Konnig:2014xva},
where subhorizon scales were analyzed in the quasistatic limit, various
modified gravity parameters were calculated, and deviations from GR
predictions were presented. It was shown that predictions of the model
are consistent with existing large-scale structure data and the model
can be tested by future experiments. IBB was, however, shown later to
suffer from two other types of instability; it violates the Higuchi
bound~\cite{Lagos:2014lca,Konnig:2015lfa}, which can potentially be dangerous as
the instabilities might appear at higher-order perturbations for nonlinear
structure, and tensor perturbations have ghost instabilities at early times~\cite{Cusin:2014psa,Amendola:2015tua,Johnson:2015tfa,Konnig:2015lfa}.
Addressing the problem of instabilities in bigravity is currently
an active field of research. In principle most of the generalizations
of the dRGT theory of massive gravity enumerated in the previous paragraph
can be applied to bigravity to investigate possible resolutions to
the instability problems and to construct viable alternatives to $\Lambda$CDM.

Motivated by these instability problems in the standard scenario of
massive bigravity with FLRW metrics, in this paper we take the first
step in exploring one potential route to resolve the obstacles and
construct a viable and stable model. It is not clear at this stage
whether possible solutions are necessarily in modifications of the
structure of the theory, i.e. at the level of the action. Such possibilities
should definitely be explored, but one should also consider cases
where the structure of the theory remains intact while other classes
of solutions are considered. One of these possibilities is the class
of solutions with non-FLRW metrics for one or both metrics of the
theory. As we mentioned earlier, this has been shown to be a promising
route in the case of dRGT, and it is, therefore, worth investigating
for bigravity as well. Before studying the cosmological implications
of such cases, one needs to check whether solutions to field equations
exist and whether they are consistent with basic constraints of the
theory, such as Bianchi constraints. This is the objective of the
present paper.

Inhomogeneous and anisotropic solutions in bigravity have been studied
in the literature. This includes Bianchi cosmologies where both metrics
are homogeneous and anisotropic~\cite{Maeda:2013bha,Kao:2014efa},
as well as cosmologies with the physical metric being FLRW and the
reference metric being inhomogeneous~\cite{Volkov:2012cf}. In this
paper we study various combinations of FLRW and non-FLRW metrics in
a systematic and more general way, and investigate for which combinations
consistent solutions to the equations of motion exist. Our aim is
to identify such combinations without further exploration of their
implications for cosmology; we leave this for future work.

The rest of this paper is organized as follows. In Sec.~\ref{sec:bigravity}
we review the Hassan-Rosen theory of singly-coupled bigravity and
present the field equations and Bianchi constraints. In Sec.~\ref{sec:metcomb}
we study various combinations of FLRW, inhomogeneous, and anisotropic
solutions for the metrics of the theory. We start this in Sec.~\ref{sec:difcurv}
with the case where both metrics are of the FLRW form but have different
spatial curvatures. Since, as we will see, this will force the metrics
to be of the Lema\^{i}tre form, we study those solutions in the same
section. We move on in Sec.~\ref{sec:frwltb} with the combinations
where one metric is FLRW and the other one is of the Lema\^{i}tre-Tolman-Bondi
(LTB) form, and in Sec.~\ref{sec:ltbltb} we investigate the solutions
where both metrics are LTB. Combinations of FLRW and anisotropic but
homogeneous metrics are studied in Sec.~\ref{sec:frwbianchi}
in the context of Bianchi type I solutions. In all these sections
we discuss the consistency of the solutions when matter sources respect
or violate the homogeneity or anisotropy assumptions. In Sec.~\ref{sec:pert}
we go beyond the background solutions and investigate the scenarios
where one metric is perturbed while the other one is kept unperturbed.
Our discussions are based on both scalar and tensor perturbations.
We discuss our results and conclude in Sec.~\ref{sec:conclusions}.

\section{The theory of massive bigravity}
\label{sec:bigravity}

The Hassan-Rosen theory of ghost-free, massive bigravity is characterized
by the action~\cite{Hassan:2011zd} 
\begin{eqnarray}
S & = & -\dfrac{M_{g}^{2}}{2}\int d^{4}x\sqrt{-\det g}R_{g}-\dfrac{M_{f}^{2}}{2}\int d^{4}x\sqrt{-\det f}R_{f}\nonumber \\
 & + & m^{4}\int d^{4}x\sqrt{-\det g}\sum_{n=0}^{4}\beta_{n}e_{n}\left(\sqrt{g^{-1}f}\right)+\int d^{4}x\sqrt{-\det g}\mathcal{L}_{m}(g,\Phi),
\end{eqnarray}
where $M_{g}$ and $M_{f}$ are Planck masses and $R_{g}$ and $R_{f}$
are the Ricci scalars for the metrics $g_{\mu\nu}$ and $f_{\mu\nu}$,
respectively. Here $g_{\mu\nu}$ is the standard, physical metric
coupled to matter fields $\Phi$ through the matter Lagrangian $\mathcal{L}_{m}$,
and $f_{\mu\nu}$ is the reference metric. The action contains five
interaction (mass) terms given in terms of five functions $e_{n}$.
These are the elementary symmetric polynomials of the eigenvalues
of the matrix $\sqrt{g^{-1}f}$, where $\sqrt{g^{-1}f}\sqrt{g^{-1}f}\equiv g^{\mu\nu}f_{\mu\nu}$.
The forms of these polynomials are presented in, e.g., Ref.~\cite{Hassan:2011zd}.
The quantities $\beta_{n}$ $(n=0,1,2,3,4)$ are free parameters of
the theory, and $m$ is the mass parameter. In the following we express
masses in units of $M_{g}^{2}$ and absorb $m^{4}$ into the parameters
$\beta_{n}$ ($m$ is not an independent parameter of the theory).
The action then becomes 
\begin{eqnarray}
S & = & -\dfrac{1}{2}\int d^{4}x\sqrt{-\det g}\, R_{g}-\dfrac{M_{f}^{2}}{2}\int d^{4}x\sqrt{-\det f}R_{f}\nonumber \\
 & + & \int d^{4}x\sqrt{-\det g}\,\sum_{n=0}^{4}\beta_{n}e_{n}\left(\sqrt{g^{-1}f}\right)+\int d^{4}x\sqrt{-\det g}\mathcal{L}_{m}(g,\Phi).\label{eq:action}
\end{eqnarray}

By varying the action (\ref{eq:action}) with respect to $g_{\mu\nu}$
one obtains the generalized Einstein equation for the physical metric,
\begin{equation}
R_{\mu\nu}^{g}-\dfrac{1}{2}g_{\mu\nu}R_{g}+\sum_{n=0}^{3}(-1)^{n}\beta_{n}g_{\mu\lambda}Y_{(n)\nu}^{\lambda}\left(\sqrt{g^{-1}f}\right)=T_{\mu\nu},\label{eq:eeg}
\end{equation}
where $R_{\mu\nu}^{g}$ is the $g$-metric Ricci tensor, and the matrices
$Y_{(n)}(X)$ are defined as~\cite{Hassan:2011zd} 
\begin{align*}
Y_{(0)}(X) & \equiv I,\\
Y_{(1)}(X) & \equiv X-I[X],\\
Y_{(2)}(X) & \equiv X^{2}-X[X]+\dfrac{1}{2}I\left([X]^{2}-[X^{2}]\right),\\
Y_{(3)}(X) & \equiv X^{3}-X^{2}[X]+\dfrac{1}{2}X\left([X]^{2}-[X^{2}]\right)-\dfrac{1}{6}I\left([X]^{3}-3[X][X^{2}]+2[X^{3}]\right),
\end{align*}
where $X\equiv\left(\sqrt{g^{-1}f}\right)$, $I$ is the identity
matrix, and $[...]$ is the trace operator.

By varying the action (\ref{eq:action}) with respect to the reference
metric $f_{\mu\nu}$ we obtain 
\begin{equation}
R_{\mu\nu}^{f}-\dfrac{1}{2}f_{\mu\nu}R_{f}+\dfrac{1}{M_{f}^{2}}\sum_{n=0}^{3}(-1)^{n}\beta_{4-n}f_{\mu\lambda}Y_{(n)\nu}^{\lambda}\left(\sqrt{g^{-1}f}\right)=0,\label{eq:eef}
\end{equation}
where $R_{\mu\nu}^{f}$ is the $f$-metric Ricci tensor. Under the
rescaling $f_{\mu\nu}\rightarrow M_{f}^{-2}f_{\mu\nu}$, the Ricci
scalar $R_{f}$ transforms as $R_{f}\rightarrow M_{f}^{2}R_{f}$, which
results in 
\begin{equation}
\sqrt{-\det f}R_{f}\rightarrow M_{f}^{-2}\sqrt{-\det f}R_{f}.
\end{equation}
The interaction terms in the action then transform as 
\begin{equation}
\sum_{n=0}^{4}\beta_{n}e_{n}\left(\sqrt{g^{-1}f}\right)\rightarrow\sum_{n=0}^{4}\beta_{n}e_{n}\left(M_{f}^{-1}\sqrt{g^{-1}f}\right).
\end{equation}
Since the elementary symmetric polynomials $e_{n}(X)$ are of order
$X^{n}$, the rescaling of $f_{\mu\nu}$ by a constant factor $M_{f}^{-2}$
translates into a redefinition of the coupling constants $\beta_{n}\rightarrow M_{f}^{n}\beta_{n}$,
which allows us to assume $M_{f}=1$.\footnote{See, however, Ref.~\cite{Akrami:2015qga}, which appeared during the completion of this work, for caveats associated with this rescaling.}

In addition to the equations of motion for the metrics, there are
additional constraints on the dynamics of the metrics coming from
the Bianchi identities and the assumption that the stress-energy-momentum
tensor of the matter components is conserved,
\begin{equation}
\frac{1}{2}\nabla_{g}^{\mu}\sum_{n=0}^{3}(-1)^{n}\beta_{n}g_{\mu\lambda}Y_{(n)\nu}^{\lambda}\left(\sqrt{g^{-1}f}\right)=0,\label{eq:bi.g}
\end{equation}
where $\nabla_{g}$ is the covariant derivative operator with respect
to $g_{\mu\nu}$. Any acceptable bigravity solution must satisfy the
generalized Einstein equations (\ref{eq:eeg}) and (\ref{eq:eef}),
as well as the Bianchi constraint (\ref{eq:bi.g}). In the rest of
this paper, we investigate various types of the physical and reference
metrics, and identify the ones which are consistent with these conditions.

\section{Consistency of background solutions with different metric combinations}
\label{sec:metcomb}

\subsection{FLRW metrics with different spatial curvatures: The need for a Lema\^{i}tre reference metric}
\label{sec:difcurv}

We begin our investigation of bigravity with nonstandard metric forms
by considering the solutions for which both metrics $g_{\mu\nu}$
and $f_{\mu\nu}$ are FLRW with generic spatial curvatures $k_{g}$
and $k_{f}$, where $k_{g},k_{f}=0,\pm1$. The usual background analysis
of the cosmology of bigravity assumes $k_{g}=k_{f}$. The reason is
partly due to the significant simplification of the calculations in
this case, and in addition it seems intuitively reasonable to assume
that the two metrics respect the same symmetries and geometries. More complicated cases, however, cannot be excluded {\it a priori}. Let
us, therefore, leave the choices for $k_{g}$ and $k_{f}$ completely
generic and study this case in terms of the consistency of cosmological solutions. Before we continue we note that this case has previously been studied in Ref.~\cite{Comelli:2011zm} and shown to be inconsistent using the Bianchi constraint. In the following, however, we use the implications of the Bianchi constraint in this case as a tool to systematically construct a particular metric combination, FLRW-Lema\^{i}tre, which is consistent. Therefore, although our results are in agreement with the findings of Ref.~\cite{Comelli:2011zm}, our approach and objectives are different.

With the assumptions made above the most general
forms for the metrics are 
\begin{align}
g_{\mu\nu}dx^{\mu}dx^{\nu} & =-dt^{2}+a^{2}\left(t\right)d\vec{x_{g}}^{2},\label{eq:g1}\\
f_{\mu\nu}dx^{\mu}dx^{\nu} & =-X^{2}\left(t\right)dt^{2}+b^{2}\left(t\right)d\vec{x_{f}}^{2},\label{eq:f1}
\end{align}
where 
\begin{align}
d\vec{x_{g}}^{2} & =\dfrac{dr^{2}}{1-k_{g}r^{2}}+r^{2}\left(d\theta^{2}+sin^{2}\left(\theta\right)d\phi^{2}\right),\\
d\vec{x_{f}}^{2} & =\dfrac{dr^{2}}{1-k_{f}r^{2}}+r^{2}\left(d\theta^{2}+sin^{2}\left(\theta\right)d\phi^{2}\right),
\end{align}
$r$, $\theta$ and $\phi$ are spherical coordinates, $a$ and $b$ are the scale factors for $g_{\mu\nu}$ and $f_{\mu\nu}$, respectively, and $X$ is the lapse for $f_{\mu\nu}$. Inserting
these metric forms into the Bianchi constraint (\ref{eq:bi.g}) yields
the condition 
\begin{equation}
X=\dfrac{\dot{b}}{\dot{a}}\dfrac{\beta_{1}\left(2+\kappa\right)+2\beta_{2}\left(1+2\kappa\right)\dfrac{b}{a}+3\beta_{3}\kappa\left(\dfrac{b}{a}\right)^{2}}{3\beta_{1}+2\beta_{2}\left(2+\kappa\right)\dfrac{b}{a}+\beta_{3}\left(1+2\kappa\right)\left(\dfrac{b}{a}\right)^{2}}\label{eq:X1}
\end{equation}
on the $f$-metric lapse $X$, where $\kappa\equiv\sqrt{\dfrac{1-k_{g}r^{2}}{1-k_{f}r^{2}}}$,
and an overdot denotes a derivative with respect to $t$. Since the
scale factors $a$ and $b$ depend only on time, it is clear from
this expression that $X$ cannot be a function of time only and is
in general a function of both $t$ and $r$ unless $k_{g}=k_{f}$.
This, therefore, shows that FLRW solutions for the two metrics with
different spatial curvatures are not allowed, already at the level
of the Bianchi constraint.

Now assuming that $X$ is a function of both $t$ and $r$, Eq. (\ref{eq:bi.g})
places a constraint on the $f$-metric scale factor $b$,
\begin{align}
\left(\beta_{3}rX'+2\left(1-\kappa\right)\left(\beta_{2}+\beta_{3}X\right)\right)b^{2}+2\left(\beta_{2}rX'+\left(1-\kappa\right)\left(\beta_{1}+\beta_{2}X\right)\right)ab+\beta_{1}rX'a^{2}=0,\label{eq:binhlaps}
\end{align}
where a prime denotes a derivative with respect to $r$. We see from this equation that for general choices of the lapse $X$,
the scale factor $b$ should also be a function of both $t$ and $r$. It can be shown that the metrics (\ref{eq:g1}) and (\ref{eq:f1}) with both $X$ and $b$ being functions of both $r$ and $t$ cannot be reformulated in FLRW forms by any coordinate transformations (see appendix \ref{app:1} for a detailed proof). We, therefore, conclude that FLRW metrics with different spatial curvatures are not consistent. 

Let us now assume that $X$ and $b$ are both functions of $r$ and
$t$. Using the Bianchi constraint (\ref{eq:bi.g}), we arrive at the expressions 
\begin{align}
X= & \dfrac{\dot{b}}{\dot{a}}\dfrac{\beta_{1}\left(2+\kappa\right)+2\beta_{2}\left(1+2\kappa\right)\dfrac{b}{a}+3\beta_{3}\kappa\left(\dfrac{b}{a}\right)^{2}}{3\beta_{1}+2\beta_{2}\left(2+\kappa\right)\dfrac{b}{a}+\beta_{3}\left(1+2\kappa\right)\left(\dfrac{b}{a}\right)^{2}},\label{eq:X2}\\
b'= & -\dfrac{\left(\beta_{3}rX'+2\left(1-\kappa\right)\left(\beta_{2}+\beta_{3}X\right)\right)b^{2}+2\left(\beta_{2}rX'+\left(1-\kappa\right)\left(\beta_{1}+\beta_{2}X\right)\right)ab+\beta_{1}rX'a^{2}}{2r\left(\left(\beta_{1}+\beta_{2}X\right)a+\left(\beta_{2}+\beta_{3}X\right)b\right)}. 
\end{align}
We can, therefore, see that because of different curvatures of reference and
physical metrics, the reference metric takes a spherically symmetric
and inhomogeneous form, where the lapse and all scale factors are
functions of both $r$ and $t$. The most general metric forms corresponding
to this case are 
\begin{align}
g_{\mu\nu}dx^{\mu}dx^{\nu} & =-dt^{2}+a^{2}\left(t\right)d\vec{x_{g}}^{2},\label{eq:lam.g}\\
f_{\mu\nu}dx^{\mu}dx^{\nu} & =-X^{2}\left(t,r\right)dt^{2}+Y^{2}\left(t,r\right)dr^{2}+Z^{2}\left(t,r\right)r^{2}d\Omega^{2},\label{eq:lam.f}
\end{align}
where $d\Omega^{2}=d\theta^{2}+\sin^{2}(\theta)d\phi^{2}$. The metric
$f_{\mu\nu}$ in Eq. (\ref{eq:lam.f}) has a generic spherically symmetric
and inhomogeneous form. In the literature~\cite{Bolejko:2008ya,Bolejko:2011jc}, this type of metric is called Lema\^{i}tre metric~\cite{Lemaitre:1933gd},
and the cosmological model built on this metric is called Lema\^{i}tre
model. In GR, the Lema\^{i}tre metric arises when we have inhomogeneous
matter sources \cite{Lasky:2010vn,Bolejko:2011jc}, in particular
when the pressure and density are functions of both temporal and spatial
coordinates. In the special case of dust or homogenous pressure the
Lema\^{i}tre metric reduces to the so-called Lema\^{i}tre-Tolman-Bondi
(LTB) metric, where the lapse does not depend on the spatial coordinates
and can be rescaled.

All our arguments so far for the metrics to take the forms (\ref{eq:lam.g})
and (\ref{eq:lam.f}) when we assume unequal curvatures were based
only on the Bianchi constraint (\ref{eq:bi.g}). We can, however, arrive
at the same conclusions by analyzing the Einstein equations. Assuming
$k_{g}\neq k_{f}$ for the metrics with FLRW forms, we have a
nonvanishing $\sqrt{(1-k_{g}r^{2})/(1-k_{f}r^{2})}$ factor in the
Einstein equations for the terms corresponding to the interactions
between the two metrics. The interaction part in the $f$-metric Einstein
equation plays the role of an inhomogeneous source for $f_{\mu\nu}$,
which forces it to take a Lema\^{i}tre form. For the $g_{\mu\nu}$
metric, there is a coupling to the matter source, and by taking an
inhomogeneous matter source one can in principle cancel the inhomogeneities
coming from the interaction terms; as a result, $g_{\mu\nu}$ can
maintain its homogenous FLRW form. This confirms our finding that
the metrics should have the forms (\ref{eq:lam.g}) and (\ref{eq:lam.f})
where the matter source is inhomogeneous.

Let us now derive the explicit forms of the Einstein equations for the metrics (\ref{eq:lam.g})
and (\ref{eq:lam.f}). As argued above, we assume that the stress-energy-momentum
tensor of the matter source coupled to the physical metric has an
inhomogeneous perfect-fluid form,
\begin{align}
T_{g0}^{0} & =-\rho\left(t,r\right),\nonumber \\
T_{g1}^{1} & =p\left(t,r\right),\nonumber \\
T_{g2}^{2} & =p\left(t,r\right),\nonumber \\
T_{g3}^{3} & =p\left(t,r\right),
\end{align}
where $\rho$ and $p$ are, respectively, the energy density and pressure for the matter source. For simplicity, here we consider only the $k_{g}=0$ case. The $g$-metric
Einstein equations for this case read 
\begin{align}
3\dfrac{\dot{a}^{2}}{a^{2}}+\rho & =\beta_{0}+\beta_{1}\dfrac{(Y+2Z)}{a}+\beta_{2}\dfrac{(2Y+Z)Z}{a^{2}}+\beta_{3}\dfrac{YZ^{2}}{a^{3}},\label{eq:syst1}\\
\dfrac{\dot{a}^{2}}{a^{2}}+2\dfrac{\ddot{a}}{a}+p & =\beta_{0}+\beta_{1}\left(X+2\dfrac{Z}{a}\right)+\beta_{2}\left(2\dfrac{XZ}{a}+\dfrac{Z^{2}}{a^{2}}\right)+\beta_{3}\dfrac{XZ^{2}}{a^{2}},\label{eq:syst2}\\
\dfrac{\dot{a}^{2}}{a^{2}}+2\dfrac{\ddot{a}}{a}+p & =\beta_{0}+\beta_{1}\left(X+\dfrac{Y+Z}{a}\right)+\beta_{2}\left(\dfrac{X(Y+Z)}{a}+\dfrac{YZ}{a^{2}}\right)+\beta_{3}\dfrac{XYZ}{a^{2}}.\label{eq:syst3}
\end{align}
The equations of motion for $f_{\mu\nu}$ are too unwieldy to be displayed
here. One can find the equations for the most general case in Ref.
\citep{Volkov:2011an}. We see from Eqs. (\ref{eq:syst1})-(\ref{eq:syst3})
that with an appropriate choice of $\rho$ and $p$ it is in principle
possible to find a function $a\left(t\right)$ which satisfies the
equations of motion.

In conclusion, our bigravity theory does not allow two FLRW metrics
with different spatial curvatures, while it is possible to have a
combination of FLRW and Lema\^{i}tre forms for the metrics $g_{\mu\nu}$
and $f_{\mu\nu}$, respectively, if the matter source takes an inhomogeneous
form. In the opposite case of a Lema\^{i}tre form for $g_{\mu\nu}$
and an FLRW form for $f_{\mu\nu}$, the equation of motion for $f_{\mu\nu}$
will contain a homogenous Einstein tensor part and an inhomogeneous
interaction part. Since $f_{\mu\nu}$ does not couple to matter, the
inhomogeneities cannot be cancelled and, therefore, this metric combination
in general does not have consistent solutions.

\subsection{FLRW--LTB and LTB--FLRW combinations}
\label{sec:frwltb}

The next possibility we wish to explore is the case where one of the
metrics is FLRW and the other one is LTB. For the case where $g_{\mu\nu}$
is FLRW and $f_{\mu\nu}$ is LTB (we denote this as FLRW--LTB), the
line elements have the forms 
\begin{align}
g_{\mu\nu}dx^{\mu}dx^{\nu} & =-dt^{2}+a^{2}(t)d\vec{x_{g}}^{2},\\
f_{\mu\nu}dx^{\mu}dx^{\nu} & =-X^{2}(t)dt^{2}+Y^{2}(t,r)dr^{2}+Z^{2}(t,r)r^{2}d\Omega^{2},
\end{align}
where again $d\Omega^{2}=d\theta^{2}+\sin^{2}(\theta)d\phi^{2}$.
Here we assume that $g_{\mu\nu}$ has a curvature $k_{g}$. In addition
we follow the standard recipe for LTB metrics and assume that the
physical metric is coupled to a homogeneous perfect-fluid source.
In this case the $(0,0)$ and $(1,1)$ components of the $g_{\mu\nu}$
equation of motion become 
\begin{equation}
-3a\dot{a}^{2}+a^{3}\left(\beta_{0}+\rho\right)+a^{2}\beta_{1}\left(2Z+Y\sqrt{1-r^{2}k_{g}}\right)+a\left(-3k_{g}+\beta_{2}Z^{2}+2\beta_{2}YZ\sqrt{1-k_{g}r^{2}}\right)+\beta_{3}YZ^{2}\sqrt{1-k_{g}r^{2}}=0,\label{eq:flrwltb1g}
\end{equation}
\begin{equation}
\dot{a}^{2}+a^{2}\left(p-\beta_{0}-\beta_{1}X\right)-2a\left(\left(\beta_{1}+\beta_{2}X\right)Z-\ddot{a}\right)+k_{g}-\beta_{2}Z^{2}-\beta_{3}XZ^{2}=0.\label{eq:flrwltb2g}
\end{equation}
Here we have assumed the stress-energy-momentum tensor for the isotropic
and homogeneous perfect fluid to be of the standard form 
\begin{equation}
T_{g\nu}^{\mu}=\left(\rho+p\right)u_{0}^{\mu}u_{0\nu}+p\delta_{\:\nu}^{\mu},\label{eq:Tghom}
\end{equation}
where $\rho=\rho(t)$ is the rest energy density of the fluid, $p=p(t)$
is its pressure, and $u_{0}^{\mu}$ is its isotropic four-velocity. It
is clear from Eq. (\ref{eq:flrwltb2g}) that in general $Z$ cannot
be a function of $r$ since all the other quantities in the equation,
including the $f$-metric lapse $X$, are functions only of $t$.
If $Z$ is a function of $t$ only, in order to satisfy Eq. (\ref{eq:flrwltb1g})
$Y$ should be of the form $Y(t,r)=A(t)/\sqrt{1-k_{g}r^{2}}$, where
$A(t)$ is an arbitrary function of $t$. This then implies that the
reference metric should also be of an FLRW type with the same curvature
$k_{g}$.

In the opposite case, where the physical metric is LTB and the reference
metric is FRLW (we denote this as LTB--FLRW), we have 
\begin{align}
g_{\mu\nu}dx^{\mu}dx^{\nu} & =-dt^{2}+Y^{2}(t,r)dr^{2}+Z^{2}(t,r)r^{2}d\Omega^{2},\\
f_{\mu\nu}dx^{\mu}dx^{\nu} & =-X^{2}(t)dt^{2}+b^{2}(t)d\vec{x_{f}}^{2}.
\end{align}
The $(0,0)$ and $(1,1)$ components of the $f_{\mu\nu}$ equation
of motion then read 
\begin{align}
X^{2}Y\left(\beta_{1}Z^{2}+2\beta_{2}Zb+\beta_{3}b^{2}\right)\sqrt{1-k_{f}r^{2}}+b\left(\beta_{2}X^{2}Z^{2}+2\beta_{3}X^{2}Zb+\beta_{4}X^{2}b^{2}-3k_{f}X^{2}-3\dot{b}^{2}\right)=0,\label{eq:flrwltb1f}\\
\beta_{4}b^{2}X^{3}-k_{f}X^{3}+\left(\beta_{1}+\beta_{2}X\right)X^{2}Z^{2}+\beta_{3}X^{2}b^{2}-X\dot{b}^{2}+2\dot{X}b\dot{b}+2bX^{2}\left(\beta_{2}+\beta_{3}X\right)Z-2Xb\ddot{b}=0.\label{eq:flrwltb2f}
\end{align}
Similarly to the case for $g_{\mu\nu}$, here we again see from Eq.
(\ref{eq:flrwltb2f}) that $Z$ in general cannot depend on $r$.
Equation (\ref{eq:flrwltb1f}), therefore, implies that $Y$ should have
the form $Y(t,r)=B(t)/\sqrt{1-k_{f}r^{2}}$, where $B\left(t\right)$
is again an arbitrary function of $t$. This form for $Y$ then brings
the physical metric into an FLRW type, again with the same curvature
as the reference metric.

It is important to note that in both metric combinations discussed
above the Bianchi constraint is not satisfied, and since we get the
Bianchi constraint using the covariant conservation of the stress-energy-momentum
tensor, we can therefore state that in general neither FLRW--LTB nor
LTB--FLRW can occur for any choices of covariantly conserved $T_{\mu\nu}^{g}$.

\subsection{LTB--LTB combination}
\label{sec:ltbltb}

Let us now study the case where both metrics $g_{\mu\nu}$ and $f_{\mu\nu}$
are of LTB forms (we denote this as LTB--LTB), 
\begin{align}
g_{\mu\nu}dx^{\mu}dx^{\nu} & =-dt^{2}+A^{2}(t,r)dr^{2}+B^{2}(t,r)d\Omega^{2},\\
f_{\mu\nu}dx^{\mu}dx^{\nu} & =-X^{2}(t)dt^{2}+Y^{2}(t,r)dr^{2}+Z^{2}(t,r)d\Omega^{2}.
\end{align}
In this case the $0$ component of the Bianchi constraint (\ref{eq:bi.g})
enforces the $f$-metric lapse $X(t)$ to satisfy the equation 
\begin{equation}
X=\dfrac{2U\dot{Z}+V\dot{Y}}{2U\dot{B}+V\dot{A}},\label{eq:zrel}
\end{equation}
where 
\begin{align}
U(t,r) & \equiv B(t,r)\left(\beta_{1}A(t,r)+\beta_{2}Y(t,r)\right)+Z(t,r)\left(\beta_{2}A(t,r)+\beta_{3}Y(t,r)\right),\\
V(t,r) & \equiv\beta_{1}B^{2}(t,r)+2\beta_{2}B(t,r)Z(t,r)+\beta_{3}Z^{2}(t,r).
\end{align}
The constraint (\ref{eq:zrel}) holds when 
\begin{align}
B(t,r) & =\widetilde{B}(t)s(r),\\
Z(t,r) & =\widetilde{Z}(t)s(r),\\
A(t,r) & =\widetilde{A}(t)q(r),\\
Y(t,r) & =\widetilde{Y}(t)q(r),
\end{align}
where $s$ and $q$ are functions of $r$ only, obtained through solving
Einstein equations. In addition, the $(1,2)$ component of the equation of motion for
$g_{\mu\nu}$ becomes 
\begin{equation}
B'\dot{A}=A\dot{B}'.
\end{equation}
For $f_{\mu\nu}$ the corresponding equation is 
\begin{equation}
Z'\dot{Y}=Y\dot{Z}'.
\end{equation}
From these equations we obtain the following relations between $\widetilde{B}(t)$,
$\widetilde{A}(t)$, $\widetilde{Z}(t)$, and $\widetilde{Y}(t)$:
\begin{align}
\widetilde{B}(t)\dot{\widetilde{A}}(t) & =\widetilde{A}(t)\dot{\widetilde{B}}(t),\\
\widetilde{Z}(t)\dot{\widetilde{Y}}(t) & =\widetilde{Y}(t)\dot{\widetilde{Z}}(t).
\end{align}
We, therefore, find that $\widetilde{A}(t)=C_{1}\widetilde{B}(t)$ and
$\widetilde{Y}(t)=C_{2}\widetilde{Z}(t)$, where $C_{1}$ and $C_{2}$
are some arbitrary constants. We obtain another useful constraint
on our functions from the $(2,2)$ component of the $g_{\mu\nu}$
equation of motion,
\begin{align}
-2B\left(\left(\beta_{1}+\beta_{2}X\right)Z-\ddot{B}\right)+\dot{B}^{2}-B^{2}\left(\beta_{0}+\beta_{1}X-p\right)-\left(\beta_{2}+\beta_{3}X\right)Z^{2}=\dfrac{B'^{2}}{A^{2}}-1.\label{eq:LTLTeqg.2}
\end{align}
It is easy to see that the $r$-dependent part of the left-hand side
of Eq. (\ref{eq:LTLTeqg.2}) is $s^{2}(r)$. The right-hand side of
Eq. (\ref{eq:LTLTeqg.2}) should also have the same dependence on
$r$ in order for the $r$ dependence of both sides of the equation
to cancel out. Therefore, in this case we have 
\begin{equation}
\dfrac{B'^{2}}{A^{2}}-1=\dfrac{s'^{2}}{q^{2}}-1=C_{3}s^{2},\label{eq:LTLTcon.1}
\end{equation}
where $C_{3}$ is another arbitrary constant. From the $(3,3)$ component
of the equation of motion for $g_{\mu\nu}$ we find 
\begin{equation}
s'q'-qs''=C_{4}sq^{3}.\label{eq:LTLTcon.2}
\end{equation}
Equations (\ref{eq:LTLTcon.1}) and (\ref{eq:LTLTcon.2}) are the conditions
which should be fulfilled by $\, s$ and $\, q$ for the consistency
of the Einstein equations for both $g_{\mu\nu}$ and $f_{\mu\nu}$.
In summary, we find that the LTB--LTB combination is consistent only
for particular subclasses of LTB metrics which satisfy all the conditions
stated above.

\subsection{Bianchi I--FLRW combination}
\label{sec:bianchifrw}

We would also like to investigate the cases where one of the two metrics
is homogeneous but anisotropic while the other one is both homogeneous
and isotropic (i.e. has an FLRW form). Here we consider
only Bianchi type I models, which are the simplest anisotropic models
and capture most of the interesting anisotropic effects. The general
properties of the cases where both metrics are anisotropic, simultaneously
diagonal, and of the same Bianchi types within the Bianchi class A,
which includes types I, II, VI$_{0}$, VII$_{0}$, VIII, and IX, are
discussed in Ref.~\cite{Maeda:2013bha}; we, therefore, do not consider
those cases in this paper. As discussed in Sec.~\ref{sec:intro},
our main motivation for studying non-FLRW solutions in bigravity is
the potential resolution of the problems with the standard scenario,
in particular the instability issues, in this framework. However, there
are also theoretical and observational arguments \cite{Harko:2014nya}
in support of an anisotropic phase in the early Universe which approached
isotropy at later times. It is, therefore, interesting also from this
perspective to see whether such solutions are allowed in bigravity
(see Ref.~\cite{Maeda:2013bha} for other motivations for studying
anisotropies in bigravity, including a potentially interesting connection
to dark matter).

Bianchi type I, or simply Bianchi I, models are spatially homogenous and flat but the expansion
rate is direction-dependent. In GR, these models have been studied
for different sources with the equation of state $p=\omega\rho$.
It has been shown in Ref.~\cite{Saha:2004mr} that for cases with
$\omega<1$ the anisotropic models evolve towards an FLRW universe,
while for $\:\omega=1$ the process of isotropization does not take
place. In the present study of Bianchi metrics we do not consider
the question of isotropization and only investigate the consistency
of such solutions in terms of the field equations and Bianchi constraint
for both isotropic and anisotropic perfect-fluid sources.

In this section we focus on the case where the physical metric is
assumed to be of an anisotropic Bianchi I form while the reference
metric is FLRW and flat; we call this case Bianchi I--FLRW. We show
that for any choices of the matter source, isotropic or anisotropic,
the Bianchi I--FLRW combination does not satisfy the conditions of
the theory and is, therefore, not a consistent solution.

The metrics for this particular case have the forms
\begin{align}
g_{\mu\nu}dx^{\mu}dx^{\nu} & =-dt^{2}+a_{1}^{2}\left(t\right)dx^{2}+a_{2}^{2}\left(t\right)dy^{2}+a_{3}^{2}\left(t\right)dz^{2},\\
f_{\mu\nu}dx^{\mu}dx^{\nu} & =-X^{2}\left(t\right)dt^{2}+b^{2}\left(t\right)d\vec{x}^{2},
\end{align}
where $d\vec{x}^{2}=dx^{2}+dy^{2}+dz^{2}$, and $x$, $y$, and $z$ are Cartesian coordinates. Here, $a_{1}$, $a_{2}$, and $a_{3}$ are the $g$-metric scale factors along different directions, and $b$ is the $f$-metric scale factor. The equation of motion
for $f_{\mu\nu}$ gives, independently of the matter source for the
$g$ metric, 
\begin{align}
X^{2}\left(\beta_{1}a_{1}a_{2}a_{3}+\beta_{2}\left(a_{1}a_{3}+a_{1}a_{2}+a_{2}a_{3}\right)b+\beta_{3}\left(a_{1}+a_{2}+a_{3}\right)b^{2}+\beta_{4}b{}^{3}\right)-3b\dot{b}^{2} & =0,\\
X^{2}\left(a_{3}\left(a_{2}\left(\beta_{1}+\beta_{2}X\right)+b\left(\beta_{2}+\beta_{3}X\right)\right)+b\left(a_{2}\left(\beta_{2}+\beta_{3}X\right)+b\left(\beta_{3}+\beta_{4}X\right)\right)\right)-X\left(2b\ddot{b}+\dot{b}{}^{2}\right)+2b\dot{X}\dot{b} & =0,\label{eq:BIFr.f2}\\
X^{2}\left(a_{3}\left(a_{1}\left(\beta_{1}+\beta_{2}X\right)+b\left(\beta_{2}+\beta_{3}X\right)\right)+b\left(a_{1}\left(\beta_{2}+\beta_{3}X\right)+b\left(\beta_{3}+\beta_{4}X\right)\right)\right)-X\left(2b\ddot{b}+\dot{b}{}^{2}\right)+2b\dot{X}\dot{b} & =0,\label{eq:BIFr.f3}\\
X^{2}\left(a_{2}\left(a_{1}\left(\beta_{1}+\beta_{2}X\right)+b\left(\beta_{2}+\beta_{3}X\right)\right)+b\left(a_{1}\left(\beta_{2}+\beta_{3}X\right)+b\left(\beta_{3}+\beta_{4}X\right)\right)\right)-X\left(2b\ddot{b}+\dot{b}{}^{2}\right)+2b\dot{X}\dot{b} & =0.\label{eq:BIFr.f4}
\end{align}
From Eqs. (\ref{eq:BIFr.f2}) and (\ref{eq:BIFr.f3}) we see that $a_{1}=a_{2}$, while Eqs. (\ref{eq:BIFr.f3}) and (\ref{eq:BIFr.f4}) imply that $a_{2}=a_{3}$; therefore $a_{1}=a_{2}=a_{3}$. This means that the physical
metric is forced to become isotropic. Therefore, using only the $f$-metric
equation of motion and no other equations or conditions of the theory, we can conclude that
one cannot have a Bianchi I physical metric while the
reference metric maintains its FLRW form. Since $f_{\mu\nu}$ is not sourced
by matter, our conclusion is general and independent of what form
the matter source takes.

\subsection{FLRW--Bianchi I combination}
\label{sec:frwbianchi}

\subsubsection{Isotropic source}

Let us now assume that the reference metric takes the Bianchi-I anisotropic
form while the physical metric is FLRW and flat; we call this case
FLRW--Bianchi I. This means that each direction has a different scale
factor in the $f$ metric. The metrics, therefore, possess the forms
\begin{align}
g_{\mu\nu}dx^{\mu}dx^{\nu} & =-dt^{2}+a^{2}\left(t\right)d\vec{x}^{2},\label{eq:iso.g}\\
f_{\mu\nu}dx^{\mu}dx^{\nu} & =-X^{2}\left(t\right)dt^{2}+b_{1}^{2}\left(t\right)dx^{2}+b_{2}^{2}\left(t\right)dy^{2}+b_{3}^{2}\left(t\right)dz^{2},\label{eq:iso.f}
\end{align}
where again $d\vec{x}^{2}=dx^{2}+dy^{2}+dz^{2}$. Here, $a$ is the scale factor for $g_{\mu\nu}$, and $b_{1}$, $b_{2}$ and $b_{3}$ are the $f$-metric scale factors along different directions. In addition, let us
assume that the matter source, which couples to the physical metric
$g_{\mu\nu}$, is a homogeneous and isotropic perfect fluid. The stress-energy-momentum
tensor has, therefore, the form given in Eq. (\ref{eq:Tghom}).

For the metric forms (\ref{eq:iso.g}) and (\ref{eq:iso.f}), the
$g$-metric Einstein equation (\ref{eq:eeg}) reads 
\begin{align}
3\dfrac{\dot{a}^{2}}{a^{2}}-\rho & =\beta_{0}+\beta_{1}\dfrac{b_{1}+b_{2}+b_{3}}{a}+\beta_{2}\dfrac{b_{1}b_{2}+b_{1}b_{3}+b_{2}b_{3}}{a^{2}}+\beta_{3}\dfrac{b_{1}b_{2}b_{3}}{a^{3}},\\
\dfrac{\dot{a}^{2}}{a^{2}}+2\dfrac{\ddot{a}}{a}+p & =\beta_{0}+\beta_{1}\left(X+\dfrac{b_{2}+b_{3}}{a}\right)+\beta_{2}\left(\dfrac{X\left(b_{2}+b_{3}\right)}{a}+\dfrac{b_{2}b_{3}}{a^{2}}\right)+\beta_{3}\dfrac{Xb_{2}b_{3}}{a^{2}},\label{eq:FrBI.f2}\\
\dfrac{\dot{a}^{2}}{a^{2}}+2\dfrac{\ddot{a}}{a}+p & =\beta_{0}+\beta_{1}\left(X+\dfrac{b_{1}+b_{3}}{a}\right)+\beta_{2}\left(\dfrac{X\left(b_{1}+b_{3}\right)}{a}+\dfrac{b_{1}b_{3}}{a^{2}}\right)+\beta_{3}\dfrac{Xb_{1}b_{3}}{a^{2}},\\
\dfrac{\dot{a}^{2}}{a^{2}}+2\dfrac{\ddot{a}}{a}+p & =\beta_{0}+\beta_{1}\left(X+\dfrac{b_{1}+b_{2}}{a}\right)+\beta_{2}\left(\dfrac{X\left(b_{1}+b_{2}\right)}{a}+\dfrac{b_{1}b_{2}}{a^{2}}\right)+\beta_{3}\dfrac{Xb_{1}b_{2}}{a^{2}}.\label{eq:FrBI.f4}
\end{align}
It is straightforward to see from these equations that $b_{1}$, $b_{2}$, and $b_{3}$
are equal. This means that if we initially assume an anisotropic form
for the reference metric while the physical metric and the matter
source are both assumed to be isotropic, the structure of the Einstein
equations automatically forces the reference metric to also be isotropic.

\subsubsection{Anisotropic source}

Let us now relax the isotropy condition on the matter source and
study the FLRW--Bianchi I scenario when the physical metric is coupled
to a source which is anisotropic and of a Bianchi I type. The stress-energy-momentum
tensor in this case takes the form 
\begin{align}
T_{g0}^{0} & =-\rho(t),\\
T_{g1}^{1} & =p_{1}(t),\\
T_{g2}^{2} & =p_{2}(t),\\
T_{g3}^{3} & =p_{3}(t),
\end{align}
where the components of the fluid pressure, $p_{1}$, $p_{2}$, and
$p_{2}$, are allowed to be different along different directions and
therefore create anisotropy in the fluid.

It follows from the Bianchi constraint (\ref{eq:bi.g}) that in this
case the $f$-metric lapse $X(t)$ is given by (here all the variables
depend only on $t$) 
\begin{align}
X=\frac{1}{\dot{a}}\dfrac{\beta_{1}a^{2}\left(b_{1}+b_{2}+b_{3}\right)\dot{}+\beta_{2}a\left(b_{1}b_{2}+b_{1}b_{3}+b_{2}b_{3}\right)\dot{}+\beta_{3}\left(b_{1}b_{2}b_{3}\right)\dot{}}{3\beta_{1}a^{2}+2\beta_{2}a\left(b_{1}+b_{2}+b_{3}\right)+\beta_{3}\left(b_{1}b_{2}+b_{1}b_{3}+b_{2}b_{3}\right)}.
\end{align}
The Einstein equation for $g_{\mu\nu}$ gives 
\begin{align}
3\dfrac{\dot{a}^{2}}{a^{2}}-\rho & =\beta_{0}+\beta_{1}\dfrac{b_{1}+b_{2}+b_{3}}{a}+\beta_{2}\dfrac{b_{1}b_{3}+b_{1}b_{2}+b_{2}b_{3}}{a^{2}}+\beta_{3}\dfrac{b_{1}b_{2}b_{3}}{a^{3}},\label{eq:FLbi.g1}\\
\dfrac{\dot{a}^{2}}{a^{2}}+2\dfrac{\ddot{a}}{a}+p_{1} & =\beta_{0}+\beta_{1}X+\left(\beta_{2}+\beta_{3}X\right)\dfrac{b_{2}b_{3}}{a^{2}}+(\beta_{1}+\beta_{2}X)\dfrac{b_{2}+b_{3}}{a},\\
\dfrac{\dot{a}^{2}}{a^{2}}+2\dfrac{\ddot{a}}{a}+p_{2} & =\beta_{0}+\beta_{1}X+\left(\beta_{2}+\beta_{3}X\right)\dfrac{b_{1}b_{3}}{a^{2}}+(\beta_{1}+\beta_{2}X)\dfrac{b_{3}+b_{1}}{a},\\
\dfrac{\dot{a}^{2}}{a^{2}}+2\dfrac{\ddot{a}}{a}+p_{3} & =\beta_{0}+\beta_{1}X+\left(\beta_{2}+\beta_{3}X\right)\dfrac{b_{1}b_{2}}{a^{2}}+(\beta_{1}+\beta_{2}X)\dfrac{b_{2}+b_{1}}{a},\label{eq:FLbi.g4}
\end{align}
and the $f$-metric Einstein equation gives 
\begin{align}
\beta_{1}a^{3}X^{2}+\beta_{2}a^{2}X^{2}\left(b_{1}+b_{2}+b_{3}\right)+\beta_{3}aX^{2}\left(b_{1}b_{2}+b_{1}b_{3}+b_{2}b_{3}\right)+\beta_{4}b_{1}b_{2}b_{3}X^{2}-b_{3}\dot{b_{1}}\dot{b_{2}}-b_{1}\dot{b_{2}}\dot{b_{3}}-b_{2}\dot{b_{1}}\dot{b_{3}} & =0,\nonumber \\
a^{2}X^{2}\left(\beta_{1}+\beta_{2}X\right)+aX^{2}\left(\beta_{2}+\beta_{3}X\right)\left(b_{2}+b_{3}\right)+b_{3}\left(\dot{X}\dot{b_{2}}+X^{2}\left(\beta_{3}+\beta_{4}X\right)b_{2}\right)+b_{2}\dot{X}\dot{b_{3}}-X\left(\dot{b_{2}}\dot{b_{3}}+b_{3}\ddot{b_{2}}+b_{2}\ddot{b_{3}}\right) & =0,\nonumber \\
a^{2}X^{2}\left(\beta_{1}+\beta_{2}X\right)+aX^{2}\left(\beta_{2}+\beta_{3}X\right)\left(b_{1}+b_{3}\right)+b_{3}\left(\dot{X}\dot{b_{1}}+X^{2}\left(\beta_{3}+\beta_{4}X\right)b_{1}\right)+b_{1}\dot{X}\dot{b_{3}}-X\left(\dot{b_{1}}\dot{b_{3}}+b_{3}\ddot{b_{1}}+b_{1}\ddot{b_{3}}\right) & =0,\nonumber \\
a^{2}X^{2}\left(\beta_{1}+\beta_{2}X\right)+aX^{2}\left(\beta_{2}+\beta_{3}X\right)\left(b_{1}+b_{2}\right)+b_{2}\left(\dot{X}\dot{b_{1}}+X^{2}\left(\beta_{3}+\beta_{4}X\right)b_{1}\right)+b_{1}\dot{X}\dot{b_{2}}-X\left(\dot{b_{1}}\dot{b_{2}}+b_{2}\ddot{b_{1}}+b_{1}\ddot{b_{2}}\right) & =0.\label{eq:FrBI.f}
\end{align}

From Eqs. (\ref{eq:FLbi.g1})-(\ref{eq:FLbi.g4}) we see that because
of the anisotropic matter source, $b_{1}$, $b_{2}$, and $b_{3}$
satisfy different algebraic equations and can, therefore, be different
from each other. Equations (\ref{eq:FrBI.f}) form a set of second-order
differential equations with respect to $b_{1}$, $b_{2}$, and $b_{3}$.
Each equation has two independent solutions and, therefore, in spite
of the fact that $b_{1}$, $b_{2}$, and $b_{3}$ satisfy the same
differential equation, in general they can be different. We can, therefore,
conclude that the combination FRLW--Bianchi I in the presence of an
anisotropic matter source is in principle possible.

\section{Consistency of solutions with only one perturbed metric}
\label{sec:pert}

In the previous sections we explored various combinations of cosmologically
interesting metric types for the two metrics of the theory of massive
bigravity. Our investigation was, however, only at the level of the
background dynamics of the Universe. In this section we study some other
interesting cases where perturbations around the background metrics
are considered.

Perturbations are clearly crucial for cosmology and it is, therefore,
interesting to ask whether there are any consistent, nonstandard
ways to perturb the metrics. In all the previous works on the perturbative
analysis of bigravity a standard recipe has been followed: the background
metrics have been assumed to be FLRW and both the physical and reference
metrics have then been perturbed around the FLRW solutions.
As stated in Sec.~\ref{sec:intro}, our main motivation in this
paper has been to try alternative solutions which might be free of
various instabilities which appear in bigravity models at the level
of perturbations. It would, therefore, be interesting if one could avoid
the instabilities by finding alternative ways of perturbing the metrics.
In addition, the usual perturbation equations are very complicated
and it would be very helpful if one could find a way to simplify the equations. One of such approaches could be to work with solutions
which leave the reference metric unperturbed while the physical metric
is perturbed as usual. This scenario is physically justified because
the reference metric does not couple to the matter components and
cannot be measured directly from observations; the reference metric
affects our observables only through its interactions with the physical
metric. In what follows we analyze the consistency of this possibility
using both scalar and tensor perturbations.

Let us start with scalar perturbations up to linear order for both
metrics. For simplicity we assume both metrics to be of the flat FLRW
type. Including only scalar perturbations and using the notations
of Ref.~\cite{Konnig:2014xva}, the line elements for the perturbed
physical and reference metrics $g_{\mu\nu}$ and $f_{\mu\nu}$ have
the forms 
\begin{align}
ds_{g}^{2} & =a^{2}\left(\eta\right)\left[-\left(1+2\Psi_{g}\right)d\eta^{2}+2\partial_{i}B_{g}dx^{i}d\eta+\left[\left(1-2\Phi_{g}\right)\delta_{ij}+2\partial_{i}\partial_{j}E_{g}\right]dx^{i}dx^{j}\right],\label{eq:pertg}\\
ds_{f}^{2} & =b^{2}\left(\eta\right)\left[-\left(1+2\Psi_{f}\right)X^{2}d\eta^{2}+2\partial_{i}B_{f}Xdx^{i}d\eta+\left[\left(1-2\Phi_{f}\right)\delta_{ij}+2\partial_{i}\partial_{j}E_{f}\right]dx^{i}dx^{j}\right].\label{eq:pertf}
\end{align}
Here $a$ and $b$ are the scale factors corresponding to $g_{\mu\nu}$
and $f_{\mu\nu}$, respectively, $X$ is the lapse for $f_{\mu\nu}$, and the perturbation quantities $\left\{\Psi_{g,f},B_{g,f},\Phi_{g,f},E_{g,f}\right\} $
are allowed to depend on both conformal time $\eta$ and space. Spatial
indices are raised and lowered by the Kronecker delta. In this section
a dot denotes a derivative with respect to the conformal time $\eta$. 

In order to write down the expression for the perturbed matter stress-energy-momentum tensor, we assume a perfect fluid with an equation of state $p=w\rho$. In addition, we describe the matter perturbations with only one scalar field $\chi$. This procedure has been proposed in Ref.~\cite{Mukhanov:1990me}. With these assumptions and conventions we have
\begin{align}
\delta T^{0}{}_{0}= & -(\rho+p)(3\Phi_{g}-E_{g,ll}-\chi_{,ll}),\nonumber \\
\delta T^{i}{}_{0}= & -(\rho+p)\dot{\chi}^{,i},\nonumber \\
\delta T^{0}{}_{i}= & (\rho+p)(B_{g,i}+\dot{\chi}_{,i}),\nonumber \\
\delta T^{i}{}_{j}= & w(\rho+p)(3\Phi_{g}-E_{g,ll}-\chi_{,ll})\delta^{i}{}_{j},
\end{align}
where we sum over indices $l$. Following Ref.~\cite{Lagos:2014lca} and after some useful gauge fixing and transformations developed in Ref.~\cite{Lagos:2013aua}, we set $\Phi_{f}=\chi=0$ and arrive at the first-order perturbation equations
\begin{align}
&2\mathcal{H}\left(3\dot{\Phi}_{g}+k^{2}\dot{E}_{g}\right)+a^{2}\left((1+w)\rho(3\Phi_{g}+k^{2}E_{g})+rZ(3\Phi_{g}+k^{2}(E_{g}-E_{f}))\right)\nonumber\\
&+2\left(k^{2}\Phi_{g}+\mathcal{H}(3\mathcal{H}\Psi_{g}-k^{2}B_{g})\right)=0,\label{G00}\\
&2(X+1)\dot{\Phi}_{g}+2\mathcal{H}(X+1)\Psi_{g}-Zr(XB_{f}-B_{g})+(1+w)\rho(1+X)B_{g}=0,\label{G0i}\\
&2(k^{2}\ddot{E}_{g}+3\ddot{\Phi}_{g})+2\mathcal{H}(3\dot{\Psi}_{g}+6\dot{\Phi}_{g}+2k^{2}\dot{E}_{g})-2k^{2}\dot{B}_{g}+3Za^{2}rX(\Psi_{f}+\Psi_{g})\nonumber \\
& +a^{2}\left(-3(1+w)\rho(2\Psi_{g}+w(3\Phi_{g}+k^{2}E_{g}))+2r(-3Z\Psi_{g}+\tilde{Z}(3\Phi_{g}+k^{2}(E_{g}-E_{f})))\right)\nonumber \\
& +2(9\mathcal{H}^{2}-k^{2})\Psi_{g}+2k^{2}(\Phi_{g}-2\mathcal{H}B_{g})=0,\label{Gii}\\
&\ddot{E}_{g}-\dot{B}_{g}+2\mathcal{H}\dot{E}_{g}+\tilde{Z}a^{2}r(E_{g}-E_{f})-\Psi_{g}-2\mathcal{H}B_{g}+\Phi_{g}=0,\label{Gij}\\
& 2r\mathcal{H}_{f}k^{2}\dot{E}_{f}-a^{2}ZX^{2}(k^{2}E_{g}-k^{2}E_{f}+3\Phi_{g})-2rX\mathcal{H}_{f}k^{2}B_{f}+6\mathcal{H}_{f}^{2}r\Psi_{f}=0,\label{F00}\\
& 2\mathcal{H}_{f}r(X+1)\Psi_{f}+Xa^{2}Z(XB_{f}-B_{g})=0,\label{F0i}\\
& rX\ddot{E}_{f}-r(-2X\mathcal{H}_{f}+\dot{X})\dot{E}_{f}-X^{2}\left(\dot{B}_{f}r+rX\Psi_{f}+2r\mathcal{H}_{f}B_{f}+a^{2}\tilde{Z}(E_{g}-E_{f})\right)=0,\label{Fij}
\end{align}
where we have defined
\begin{align}
Z & \equiv\beta_{1}+2\beta_{2}r+\beta_{3}r^{2}\\
\tilde{Z} & \equiv\beta_{1}+\beta_{2}r\left(1+X\right)+\beta_{3}r^{2}X,\\
r & \equiv\dfrac{b}{a},\quad\mathcal{H}\equiv\frac{\dot{a}}{a}\quad\mathcal{H}_{f}\equiv\frac{\dot{b}}{b}.
\end{align}
Note that $r$ in this section is not the radial coordinate. Let us now assume that only the physical metric $g_{\mu\nu}$ is perturbed; i.e., all
the perturbative quantities for $f_{\mu\nu}$, $\left\{\Psi_{f},B_{f},\Phi_{f},E_{f}\right\}$,
are vanishing. Looking at Eqs. (\ref{F00})-(\ref{Fij}) we find
\begin{align}
X^{2}a^{2}Z(k^{2}E_{g}+3\Phi_{g}) & =0,\label{eq:cn.1}\\
Xa^{2}ZB_{g} & =0,\label{eq:cn.2}\\
X^{2}a^{2}\tilde{Z}E_{g} & =0.\label{eq:cn.3}
\end{align}
From Eqs. (\ref{eq:cn.1})-(\ref{eq:cn.3}) it is clear that the
$g$-metric scalar perturbations, $B_{g}$, $\Phi_{g}$ and $E_{g}$, should vanish (which in turn implies, using Eq. (\ref{Gij}), that $\Psi_{g}$ should also be vanishing) unless
the quantities $\tilde{Z}$ and $Z$ are both vanishing. In order to prove
that the latter cannot be the case let us now assume $\tilde{Z}=0$.
In this case we have 
\begin{equation}
\beta_{1}+\beta_{2}r\left(1+X\right)+\beta_{3}r^{2}X=0.\label{eq:Z0}
\end{equation}
On the other hand, we know from the Bianchi constraint (\ref{eq:bi.g}) that at the background level $X$ satisfies
the condition 
\begin{equation}
X=\dfrac{a}{\dot{a}}\dfrac{\dot{b}}{b}.\label{eq:Xconf}
\end{equation}
If we now insert the value of $X$ from Eq. (\ref{eq:Xconf}) into
Eq. (\ref{eq:Z0}) we obtain 
\begin{equation}
\beta_{1}a\dot{a}+\beta_{2}(a\dot{b}+b\dot{a})+\beta_{3}b\dot{b}=0\quad\Longrightarrow\quad\dfrac{\beta_{1}}{2}\left(a^{2}\right)\dot{}+\beta_{2}\left(ab\right)\dot{}+\dfrac{\beta_{3}}{2}\left(b^{2}\right)\dot{}=0.\label{eq:witder2-1}
\end{equation}
Finally, by integrating Eq. (\ref{eq:witder2-1}) over conformal time
we get 
\begin{equation}
\beta_{1}a^{2}+2\beta_{2}ab+\beta_{3}b^{2}=C\quad\Longrightarrow\quad\beta_{1}+2\beta_{2}r+\beta_{3}r^{2}=\frac{C}{a^2},\label{eq:witder3-1}
\end{equation}
where $C$ is an arbitrary constant. We, however, need to assume $Z=0$ in this case, which then
implies $C=0$. The condition (\ref{eq:witder3-1})
with $C=0$ requires $r$ to be a constant and given by a particular combination of the parameters
$\beta_{1}$, $\beta_{2}$ and $\beta_{3}$. From
the background equations of bimetric gravity~\cite{Konnig:2014xva} one realizes immediately
that this condition implies that the Universe is in a de Sitter state
(or, trivially, that $a=b=0$). 

Let us now study the opposite scenario, i.e. where only the reference metric $f_{\mu\nu}$ is perturbed and the physical metric remains unperturbed. In this case the $g$-metric perturbations $\left\{\Psi_{g},B_{g},\Phi_{g},E_{g}\right\}$ are vanishing. Now Eqs. (\ref{G00})-(\ref{Gij}) imply
\begin{align}
Zra^2k^2E_{f} & =0,\label{eq:cnf.1}\\
ZrXB_{f} & =0,\label{eq:cnf.2}\\
3Za^2rX\Psi_{f}-2\tilde{Z}a^2rk^2E_{f} & =0\label{eq:cnf.3}\\
\tilde{Z}a^2rE_{f} & =0.\label{eq:cnf.4}
\end{align}
From Eqs. (\ref{eq:cnf.1})-(\ref{eq:cnf.4}) we can again conclude that the
$f$-metric scalar perturbations should also vanish in this case (note that $\Phi_{f}$ is already vanishing due to our gauge choice), unless
both $\tilde{Z}$ and $Z$ are vanishing. The latter case again corresponds to a de Sitter universe where $r=\mathrm{const.}$ We can, therefore, conclude, based on both cases studied here, that for any cosmological configurations different from a pure de Sitter universe,
the possibility that only one of the two metrics is perturbed is excluded at the
scalar level.

A similar analysis can be done using tensor perturbations. The line
elements for the physical and reference metrics in this case take
the forms 
\begin{align}
ds_{g}^{2} & =a^{2}\left(\eta\right)\left[-d\eta^{2}+\left(\delta_{ij}+h_{ij}^{g}\right)dx^{i}dx^{j}\right],\\
ds_{f}^{2} & =b^{2}\left(\eta\right)\left[-X^{2}d\eta^{2}+\left(\delta_{ij}+h_{ij}^{f}\right)dx^{i}dx^{j}\right],
\end{align}
where $h_{ij}^{g,f}$ are the tensor perturbation quantities and are in general
functions of space and conformal time $\eta$. These quantities satisfy
the relations 
\begin{equation}
{h}_{i}^{i}=0,\qquad h_{ij}^{,i}=0,\label{eq:ten.p.con}
\end{equation}
for both $h^{g}$ and $h^{f}$; here we again use the Kronecker delta
to raise or lower spatial indices.

In a perfect-fluid model there are no tensor perturbation modes in the stress-energy-momentum tensor, and
as a result, the right-hand side of the perturbed Einstein equation
for $g_{\mu\nu}$ vanishes. Because of the conditions (\ref{eq:ten.p.con})
we have only two degrees of freedom for each $h_{ij}$ which are fully
decoupled and satisfy the following first-order perturbation equations
in Fourier space: 
\begin{align}
\ddot{h^{g}}+2\mathcal{H}\dot{h^{g}}+k^{2}h^{g}+a^{2}r\tilde{Z}\left(h^{g}-h^{f}\right) & =0,\label{eq:perttensg}\\
\ddot{h^{f}}+\left(2\mathcal{H}_{f}-\dfrac{\dot{X}}{X}\right)\dot{h^{f}}+k^{2}X^{2}h^{f}+\dfrac{a^{2}X}{r}\tilde{Z}\left(h^{f}-h^{g}\right) & =0.\label{eq:perttensf}
\end{align}

As in the scalar case discussed above, we now assume that only the $g$
metric is perturbed while the $f$ metric is unperturbed (the argument
is identical if we assume instead that only $f_{\mu\nu}$ is perturbed); this means that $h^{f}=0$. From Eq. (\ref{eq:perttensf}) it is straightforward
to see that the $g$-metric tensor perturbation $h^{g}$ should
also vanish, unless the quantity $\tilde{Z}$ is vanishing. This corresponds
to the condition (\ref{eq:witder3-1}), but contrary to the scalar
case, here $C$ is an arbitrary constant and is not required to vanish in order for the $g$ and $f$ tensor perturbations to decouple. One obvious solution is, however, again the case where $C=0$, i.e. a purely de Sitter universe which does not correspond to the real universe. The only nontrivial class of solutions for which the tensor modes decouple while the scalar modes are still coupled is the case where $C$ is nonvanishing. Whether or not these solutions are cosmologically interesting remains to be investigated; we leave this for future work.\footnote{The case for vector perturbations is not of interest to us here, but looking at the corresponding perturbative equations presented in Ref.~\cite{Lagos:2013aua} it seems to us that the case where vector perturbations are nonvanishing only for the physical metric while the reference metric remains unperturbed is possible.} 

\section{Conclusions}
\label{sec:conclusions}

\begin{table}[t]
\begin{centering}
\begin{tabular}{|c|c|c|c|c|}
\hline 
\multicolumn{5}{|c|}{Metric combinations}\tabularnewline
\hline 
$g_{\mu\nu}$ (physical metric) & $f_{\mu\nu}$ (reference metric) & $T_{\mu\nu}^{g}$ & Possibility & Reference \tabularnewline
\hline 
\hline 
FLRW ($k$) & FLRW ($k$) & PF & {\large $\checkmark$} & Standard \tabularnewline
\hline
FLRW ($k_{g}$) & FLRW ($k_{f}$) & PF & {\large $\times$} & Present work and Ref.~\cite{Comelli:2011zm} \tabularnewline
\hline
FLRW  & Lema\^{i}tre  & PF & {\large $\times$} & Present work \tabularnewline
\hline
FLRW  & Lema\^{i}tre  & Inhom. & {\large $\checkmark$} & Present work and Ref.~\cite{Volkov:2012cf} \tabularnewline
\hline 
Lema\^{i}tre  & FLRW  & Any & {\large $\times$} & Present work \tabularnewline
\hline
FLRW  & LTB & Any & {\large $\times$} & Present work \tabularnewline
\hline 
LTB  & FLRW & Any & {\large $\times$} & Present work \tabularnewline
\hline
LTB  & LTB & PF & {\large $\checkmark$}\footnote{There are conditions which must be satisfied; see the text for details.} & Present work \tabularnewline
\hline
Bianchi I  & FLRW & Any & {\large $\times$} & Present work \tabularnewline
\hline
FLRW  & Bianchi I & PF & {\large $\times$} & Present work \tabularnewline
\hline
FLRW  & Bianchi I & Aniso. & {\large $\checkmark$} & Present work \tabularnewline
\hline
Bianchi Class A & Bianchi Class A & PF & {\large $\checkmark$} & Refs.~\cite{Maeda:2013bha,Kao:2014efa} \tabularnewline
\hline
Perturbed FLRW & Perturbed FLRW & Perturbed PF & {\large $\checkmark$} & Standard \tabularnewline
\hline
Perturbed FLRW (scalars) & Unperturbed FLRW (scalars) & Perturbed PF & {\large $\times$}\footnote{Unless the metrics are de Sitter; see the text for details.}& Present work \tabularnewline
\hline
Unperturbed FLRW (scalars) & Perturbed FLRW (scalars) & Perturbed PF & {\large $\times$}\textsuperscript{b}& Present work \tabularnewline
\hline
Perturbed FLRW (tensors) & Unperturbed FLRW (tensors) & Perturbed PF & {\large $\times$}\footnote{Unless the metrics are de Sitter or satisfy a specific condition; see the text for details.} & Present work \tabularnewline
\hline
Unperturbed FLRW (tensors) & Perturbed FLRW (tensors) & Perturbed PF & {\large $\times$}\textsuperscript{c} & Present work \tabularnewline
\hline
\end{tabular}
\par\end{centering}
\caption{\label{tab:List-of-all}List of all metric combinations studied in the present work, as well as a few other interesting combinations studied in the literature.
$\checkmark$ and $\times$ denote consistent and inconsistent cases, respectively. Here, ``PF," ``Inhom." and ``Aniso." stand, respectively, for isotropic and homogeneous (perfect fluid),
isotropic but inhomogeneous, and homogeneous but anisotropic matter sources. ``Any" stands for a matter source of any type.
$k$ is the spatial curvature of an FLRW metric; in cases with explicit indices $g$ and $f$ for $k$ the two metrics are assumed to have different spatial curvatures.}
\end{table}

One of the problems of contemporary cosmology is that there are very
few, if any, well-studied, and well-motivated alternatives to the standard model that are
viable and distinguishable from $\Lambda$CDM. Massive bigravity models (e.g. the particular model discussed in \cite{Konnig:2014xva})
could be one of such rare cases if the problem of perturbation instability (\cite{Lagos:2014lca,Cusin:2014psa}) could be overcome. However, most
studies of bigravity models confined themselves to homogeneous and
isotropic metrics with no or identical spatial
curvatures. Before ruling out bigravity cosmology one should, therefore,
see if the problems could be solved or alleviated when a different
geometry is chosen. This paper was devoted to addressing this question
by scanning several relatively simple possible metric combinations
in search for consistent cases. Future work will be necessary to actually
solve the equations for some of such cases and identify cosmologically viable ones.

We, however, found that there are only a few metric combinations that survive our
analysis. In most cases, two different metrics are impossible
unless a suitable modification is made to the matter source beyond the standard
perfect-fluid assumption. Our results are summarized in Table \ref{tab:List-of-all}.
In particular we find that the only alternative combination with standard matter
source is an LTB--LTB model (in addition to the Bianchi-Bianchi models previously studied in Refs.~\cite{Maeda:2013bha,Kao:2014efa}), subject to some constraints. Combinations
like FLRW--Lema\^{i}tre and FLRW--Bianchi I all require either
inhomogeneous or anisotropic sources. We also investigated the question
of whether having linear perturbations in just
one metric is theoretically consistent. We found, not unexpectedly, that there are
no consistent cases except for purely de Sitter backgrounds.

Whether any of the surviving combinations give rise to viable cosmological
models at both the background and perturbative levels is, however,
entirely to be seen.

\acknowledgments We thank Frank K\"onnig and Miguel Zumalac\'arregui for helpful discussions. We also acknowledge support from DFG through the project
TRR33 ``The Dark Universe.'' 

\appendix

\section{Impossibility of mapping Lema\^{i}tre to FLRW in bigravity}
\label{app:1}

In this appendix we prove that it is impossible to map the case where the physical metric $g_{\mu\nu}$ is FLRW and the reference metric $f_{\mu\nu}$ has a lapse and scale factor depending on both $r$ and $t$ to a combination where both metrics are of an FRLW form.

We start with the metrics in terms of the coordinates $r$ and $t$,
\begin{align}
g_{\mu\nu}dx^{\mu}dx^{\nu} & =-dt^{2}+a^{2}\left(t\right)d\vec{x_{g}}^{2},\label{eq:appg1-1}\\
f_{\mu\nu}dx^{\mu}dx^{\nu} & =-X^{2}\left(t,r\right)dt^{2}+b^{2}\left(t,r\right)d\vec{x_{f}}^{2},\label{eq:f1-1}
\end{align}
where 
\begin{align}
d\vec{x_{g}}^{2} & =\dfrac{dr^{2}}{1-k_{g}r^{2}}+r^{2}d\Omega^{2},\label{eq:gmetric}\\
d\vec{x_{f}}^{2} & =\dfrac{dr^{2}}{1-k_{f}r^{2}}+r^{2}d\Omega^{2},\label{eq:fmetric}
\end{align}
and $d\Omega^{2}=d\theta^{2}+sin^{2}\left(\theta\right)d\phi^{2}$. We now want to see if it is possible under any conditions to rewrite the metrics in an FLRW form when we transform the coordinates $r$ and $t$ to some new coordinates $\tilde{r}$ and $\tilde{t}$. We assume the new and old coordinates to be related as
\begin{align}
t & =T\left(\tilde{t},\tilde{r}\right),\\
r & =R\left(\tilde{t},\tilde{r}\right).
\end{align}
Under these transformations the $g$-metric scale factor $a$ and the $f$-metric scale factor and lapse, $b$ and $X$, should transform as
\begin{align}
a\left(t\right) & \rightarrow\tilde{a}\left(\tilde{t},\tilde{r}\right),\\
X\left(t,r\right) & \rightarrow\tilde{X}\left(\tilde{t},\tilde{r}\right),\\
b\left(t,r\right) & \rightarrow\tilde{b}\left(\tilde{t},\tilde{r}\right).
\end{align}
In order to know how the line elements (\ref{eq:gmetric}) and (\ref{eq:fmetric}) look under the coordinate transformations, we should first see how $dt^{2}$ and $dr^{2}$ transform. We have
\begin{align}
dt^{2} & =\dot{T}^{2}d\tilde{t}^{2}+2T'\dot{T}d\tilde{t}d\tilde{r}+T'^{2}d\tilde{r}^{2},\\
dr^{2} & =\dot{R}^{2}d\tilde{t}^{2}+2R'\dot{R}d\tilde{t}d\tilde{r}+R'^{2}d\tilde{r}^{2},
\end{align}
where an overdot denotes a derivative with respect to $\tilde{t}$ and a prime denotes a derivative with respect to $\tilde{r}$. Using all these relations, the transformed line elements for the metrics read
\begin{align}
d\tilde{s}_{g}^{2} & =-\left(\dot{T}^{2}-\dfrac{\tilde{a}^{2}}{1-k_{g}R^{2}}\dot{R}^{2}\right)d\tilde{t}^{2}+2\left(\dfrac{\tilde{a}^{2}}{1-k_{R}R^{2}}R'\dot{R}-T'\dot{T}\right)d\tilde{t}d\tilde{r}+\left(\dfrac{\tilde{a}^{2}}{1-k_{g}R^{2}}R'^{2}-R'^{2}\right)d\tilde{r}{}^{2}+\tilde{a}^{2}R^{2}d\Omega^{2},\label{eq:transmetricg}\\
d\tilde{s}_{f}^{2} & =-\left(\tilde{X}^{2}\dot{T}^{2}-\dfrac{\tilde{b}^{2}}{1-k_{f}R^{2}}\dot{R}^{2}\right)d\tilde{t}^{2}+2\left(\dfrac{\tilde{b}^{2}}{1-k_{f}R^{2}}R'\dot{R}-\tilde{X}^{2}T'\dot{T}\right)d\tilde{t}d\tilde{r}+\left(\dfrac{\tilde{b}^{2}}{1-k_{f}R^{2}}R'^{2}-\tilde{X}^{2}T'^{2}\right)d\tilde{r}^{2}+\tilde{b}^{2}R^{2}d\Omega^{2}.\label{eq:transmetricf}
\end{align}

Now in order to have both metrics in an FLRW form, we need to set the
following constraints on our transformed metric components. From Eq. (\ref{eq:transmetricg}) we obtain
\begin{align}
\dot{T}^{2}-\dfrac{\tilde{a}^{2}}{1-k_{g}R^{2}}\dot{R}^{2} & =A^{2}\left(\tilde{t}\right),\label{eq:appg1}\\
\dfrac{\tilde{a}^{2}}{1-k_{g}R^{2}}R'\dot{R}-T'\dot{T} & =0,\label{eq:g2}\\
\dfrac{\tilde{a}^{2}}{1-k_{g}R^{2}}R'^{2}-T'^{2} & =\dfrac{B^{2}(\tilde{t})}{1-\tilde{k}_{g}\tilde{r}^{2}},\label{eq:g3}\\
\tilde{a}^{2}R^{2} & =B^{2}(\tilde{t})\tilde{r}^{2},\label{eq:g4}
\end{align}
and from Eq. (\ref{eq:transmetricf}) we find
\begin{align}
\tilde{X}^{2}\dot{T}^{2}-\dfrac{\tilde{b}^{2}}{1-k_{f}R^{2}}\dot{R}^{2} & =C^{2}\left(\tilde{t}\right),\label{eq:appf1}\\
\dfrac{\tilde{b}^{2}}{1-k_{f}R^{2}}R'\dot{R}-\tilde{X}^{2}T'\dot{T} & =0,\label{eq:f2}\\
\dfrac{\tilde{b}^{2}}{1-k_{f}R^{2}}R'^{2}-\tilde{X}^{2}T'^{2} & =\dfrac{D^{2}(\tilde{t})}{1-\tilde{k}_{f}\tilde{r}^{2}},\label{eq:f3}\\
\tilde{b}^{2}R^{2} & =D^{2}(\tilde{t})\tilde{r}^{2},\label{eq:f4}
\end{align}
where $A$, $B$, $C$, and $D$ are arbitrary functions of only $\tilde{t}$.
Now from Eqs. (\ref{eq:g4}) and (\ref{eq:f4}), we get 
\begin{equation}
\dfrac{\tilde{a}^{2}}{\tilde{b}^{2}}=\dfrac{B^{2}(\tilde{t})}{D^{2}(\tilde{t})},\label{eq:appcn.1}
\end{equation}
and from Eqs. (\ref{eq:g2}) and (\ref{eq:f2}), we get 
\begin{equation}
\dfrac{\widetilde{b}^{2}}{\left(1-k_{f}R^{2}\right)\widetilde{X}^{2}}=\dfrac{\widetilde{a}^{2}}{1-k_{g}R^{2}}.\label{eq:cn.d2}
\end{equation}
Now using the condition (\ref{eq:cn.d2}) and Eqs. (\ref{eq:appg1})
and (\ref{eq:appf1}), we find 
\begin{equation}
A^{2}\left(\tilde{t}\right)=\dfrac{C^{2}\left(\tilde{t}\right)}{\tilde{X}^{2}}.\label{eq:appcn.3}
\end{equation}
This tells us that $\widetilde{X}$ must be a function only of $\tilde{t}$. Combining this with Eqs. (\ref{eq:appcn.1}) and (\ref{eq:cn.d2}),
we immediately see that $R$ must also be a function only of $\tilde{t}$,
and as a result $R'=0$. Taking this into account, Eq. (\ref{eq:f2})
implies that 
\begin{equation}
T'\dot{T}=0.
\end{equation}
Let us first discuss the option where $\dot{T}=0$. In this case
Eq. (\ref{eq:appg1}) tells us that $\tilde{a}$ must be a function only
of $\tilde{t}$. Given that $\tilde{a}$ and $\tilde{R}$ are both functions only of $\tilde{t}$, Eq. (\ref{eq:g4})
immediately gives us a contradiction since the left-hand side is a function only of $\tilde{t}$, while the right-hand side is a function of both $\tilde{t}$ and $\tilde{r}$. For the second option where we assume $T'=0$, Eqs.
(\ref{eq:g3}) and (\ref{eq:f3}) imply that $D(\tilde{t})=B(\tilde{t})=0$,
which of course means that we cannot have FRLW metrics under any coordinate transformations.

\bibliography{bibliography,amendola}

%merlin.mbs apsrev4-1.bst 2010-07-25 4.21a (PWD, AO, DPC) hacked
%Control: key (0)
%Control: author (8) initials jnrlst
%Control: editor formatted (1) identically to author
%Control: production of article title (-1) disabled
%Control: page (0) single
%Control: year (1) truncated
%Control: production of eprint (0) enabled
\begin{thebibliography}{88}%
\makeatletter
\providecommand \@ifxundefined [1]{%
 \@ifx{#1\undefined}
}%
\providecommand \@ifnum [1]{%
 \ifnum #1\expandafter \@firstoftwo
 \else \expandafter \@secondoftwo
 \fi
}%
\providecommand \@ifx [1]{%
 \ifx #1\expandafter \@firstoftwo
 \else \expandafter \@secondoftwo
 \fi
}%
\providecommand \natexlab [1]{#1}%
\providecommand \enquote  [1]{``#1''}%
\providecommand \bibnamefont  [1]{#1}%
\providecommand \bibfnamefont [1]{#1}%
\providecommand \citenamefont [1]{#1}%
\providecommand \href@noop [0]{\@secondoftwo}%
\providecommand \href [0]{\begingroup \@sanitize@url \@href}%
\providecommand \@href[1]{\@@startlink{#1}\@@href}%
\providecommand \@@href[1]{\endgroup#1\@@endlink}%
\providecommand \@sanitize@url [0]{\catcode `\\12\catcode `\$12\catcode
  `\&12\catcode `\#12\catcode `\^12\catcode `\_12\catcode `\%12\relax}%
\providecommand \@@startlink[1]{}%
\providecommand \@@endlink[0]{}%
\providecommand \url  [0]{\begingroup\@sanitize@url \@url }%
\providecommand \@url [1]{\endgroup\@href {#1}{\urlprefix }}%
\providecommand \urlprefix  [0]{URL }%
\providecommand \Eprint [0]{\href }%
\providecommand \doibase [0]{http://dx.doi.org/}%
\providecommand \selectlanguage [0]{\@gobble}%
\providecommand \bibinfo  [0]{\@secondoftwo}%
\providecommand \bibfield  [0]{\@secondoftwo}%
\providecommand \translation [1]{[#1]}%
\providecommand \BibitemOpen [0]{}%
\providecommand \bibitemStop [0]{}%
\providecommand \bibitemNoStop [0]{.\EOS\space}%
\providecommand \EOS [0]{\spacefactor3000\relax}%
\providecommand \BibitemShut  [1]{\csname bibitem#1\endcsname}%
\let\auto@bib@innerbib\@empty
%</preamble>
\bibitem [{\citenamefont {Martin}(2012)}]{Martin:2012bt}%
  \BibitemOpen
  \bibfield  {author} {\bibinfo {author} {\bibfnamefont {J.}~\bibnamefont
  {Martin}},\ }\href {\doibase 10.1016/j.crhy.2012.04.008} {\bibfield
  {journal} {\bibinfo  {journal} {Comptes Rendus Physique}\ }\textbf {\bibinfo
  {volume} {13}},\ \bibinfo {pages} {566} (\bibinfo {year} {2012})},\ \Eprint
  {http://arxiv.org/abs/1205.3365} {arXiv:1205.3365 [astro-ph.CO]} \BibitemShut
  {NoStop}%
%%CITATION = ARXIV:1205.3365;%%
\bibitem [{\citenamefont {{Amendola}}\ and\ \citenamefont
  {{Tsujikawa}}(2010)}]{2010deto.book.....A}%
  \BibitemOpen
  \bibfield  {author} {\bibinfo {author} {\bibfnamefont {L.}~\bibnamefont
  {{Amendola}}}\ and\ \bibinfo {author} {\bibfnamefont {S.}~\bibnamefont
  {{Tsujikawa}}},\ }\href@noop {} {\emph {\bibinfo {title} {{Dark Energy:
  Theory and Observations}}}}\ (\bibinfo {year} {2010})\BibitemShut {NoStop}%
\bibitem [{\citenamefont {Clifton}\ \emph {et~al.}(2012)\citenamefont
  {Clifton}, \citenamefont {Ferreira}, \citenamefont {Padilla},\ and\
  \citenamefont {Skordis}}]{Clifton:2011jh}%
  \BibitemOpen
  \bibfield  {author} {\bibinfo {author} {\bibfnamefont {T.}~\bibnamefont
  {Clifton}}, \bibinfo {author} {\bibfnamefont {P.~G.}\ \bibnamefont
  {Ferreira}}, \bibinfo {author} {\bibfnamefont {A.}~\bibnamefont {Padilla}}, \
  and\ \bibinfo {author} {\bibfnamefont {C.}~\bibnamefont {Skordis}},\ }\href
  {\doibase 10.1016/j.physrep.2012.01.001} {\bibfield  {journal} {\bibinfo
  {journal} {Phys.Rept.}\ }\textbf {\bibinfo {volume} {513}},\ \bibinfo {pages}
  {1} (\bibinfo {year} {2012})},\ \Eprint {http://arxiv.org/abs/1106.2476}
  {arXiv:1106.2476 [astro-ph.CO]} \BibitemShut {NoStop}%
%%CITATION = ARXIV:1106.2476;%%
\bibitem [{\citenamefont {Fierz}\ and\ \citenamefont
  {Pauli}(1939)}]{Fierz:1939ix}%
  \BibitemOpen
  \bibfield  {author} {\bibinfo {author} {\bibfnamefont {M.}~\bibnamefont
  {Fierz}}\ and\ \bibinfo {author} {\bibfnamefont {W.}~\bibnamefont {Pauli}},\
  }\href {\doibase 10.1098/rspa.1939.0140} {\bibfield  {journal} {\bibinfo
  {journal} {Proc.Roy.Soc.Lond.}\ }\textbf {\bibinfo {volume} {A173}},\
  \bibinfo {pages} {211} (\bibinfo {year} {1939})}\BibitemShut {NoStop}%
%%CITATION = PRSLA,A173,211;%%
\bibitem [{\citenamefont {Boulware}\ and\ \citenamefont
  {Deser}(1972)}]{Boulware:1973my}%
  \BibitemOpen
  \bibfield  {author} {\bibinfo {author} {\bibfnamefont {D.}~\bibnamefont
  {Boulware}}\ and\ \bibinfo {author} {\bibfnamefont {S.}~\bibnamefont
  {Deser}},\ }\href {\doibase 10.1103/PhysRevD.6.3368} {\bibfield  {journal}
  {\bibinfo  {journal} {Phys.Rev.}\ }\textbf {\bibinfo {volume} {D6}},\
  \bibinfo {pages} {3368} (\bibinfo {year} {1972})}\BibitemShut {NoStop}%
%%CITATION = PHRVA,D6,3368;%%
\bibitem [{\citenamefont {de~Rham}\ and\ \citenamefont
  {Gabadadze}(2010)}]{deRham:2010ik}%
  \BibitemOpen
  \bibfield  {author} {\bibinfo {author} {\bibfnamefont {C.}~\bibnamefont
  {de~Rham}}\ and\ \bibinfo {author} {\bibfnamefont {G.}~\bibnamefont
  {Gabadadze}},\ }\href {\doibase 10.1103/PhysRevD.82.044020} {\bibfield
  {journal} {\bibinfo  {journal} {Phys.Rev.}\ }\textbf {\bibinfo {volume}
  {D82}},\ \bibinfo {pages} {044020} (\bibinfo {year} {2010})},\ \Eprint
  {http://arxiv.org/abs/1007.0443} {arXiv:1007.0443 [hep-th]} \BibitemShut
  {NoStop}%
%%CITATION = ARXIV:1007.0443;%%
\bibitem [{\citenamefont {de~Rham}\ \emph
  {et~al.}(2011{\natexlab{a}})\citenamefont {de~Rham}, \citenamefont
  {Gabadadze},\ and\ \citenamefont {Tolley}}]{deRham:2010kj}%
  \BibitemOpen
  \bibfield  {author} {\bibinfo {author} {\bibfnamefont {C.}~\bibnamefont
  {de~Rham}}, \bibinfo {author} {\bibfnamefont {G.}~\bibnamefont {Gabadadze}},
  \ and\ \bibinfo {author} {\bibfnamefont {A.~J.}\ \bibnamefont {Tolley}},\
  }\href {\doibase 10.1103/PhysRevLett.106.231101} {\bibfield  {journal}
  {\bibinfo  {journal} {Phys.Rev.Lett.}\ }\textbf {\bibinfo {volume} {106}},\
  \bibinfo {pages} {231101} (\bibinfo {year} {2011}{\natexlab{a}})},\ \Eprint
  {http://arxiv.org/abs/1011.1232} {arXiv:1011.1232 [hep-th]} \BibitemShut
  {NoStop}%
%%CITATION = ARXIV:1011.1232;%%
\bibitem [{\citenamefont {de~Rham}\ \emph {et~al.}(2012)\citenamefont
  {de~Rham}, \citenamefont {Gabadadze},\ and\ \citenamefont
  {Tolley}}]{deRham:2011rn}%
  \BibitemOpen
  \bibfield  {author} {\bibinfo {author} {\bibfnamefont {C.}~\bibnamefont
  {de~Rham}}, \bibinfo {author} {\bibfnamefont {G.}~\bibnamefont {Gabadadze}},
  \ and\ \bibinfo {author} {\bibfnamefont {A.~J.}\ \bibnamefont {Tolley}},\
  }\href {\doibase 10.1016/j.physletb.2012.03.081} {\bibfield  {journal}
  {\bibinfo  {journal} {Phys.Lett.}\ }\textbf {\bibinfo {volume} {B711}},\
  \bibinfo {pages} {190} (\bibinfo {year} {2012})},\ \Eprint
  {http://arxiv.org/abs/1107.3820} {arXiv:1107.3820 [hep-th]} \BibitemShut
  {NoStop}%
%%CITATION = ARXIV:1107.3820;%%
\bibitem [{\citenamefont {de~Rham}\ \emph
  {et~al.}(2011{\natexlab{b}})\citenamefont {de~Rham}, \citenamefont
  {Gabadadze},\ and\ \citenamefont {Tolley}}]{deRham:2011qq}%
  \BibitemOpen
  \bibfield  {author} {\bibinfo {author} {\bibfnamefont {C.}~\bibnamefont
  {de~Rham}}, \bibinfo {author} {\bibfnamefont {G.}~\bibnamefont {Gabadadze}},
  \ and\ \bibinfo {author} {\bibfnamefont {A.~J.}\ \bibnamefont {Tolley}},\
  }\href {\doibase 10.1007/JHEP11(2011)093} {\bibfield  {journal} {\bibinfo
  {journal} {JHEP}\ }\textbf {\bibinfo {volume} {1111}},\ \bibinfo {pages}
  {093} (\bibinfo {year} {2011}{\natexlab{b}})},\ \Eprint
  {http://arxiv.org/abs/1108.4521} {arXiv:1108.4521 [hep-th]} \BibitemShut
  {NoStop}%
%%CITATION = ARXIV:1108.4521;%%
\bibitem [{\citenamefont {Hassan}\ and\ \citenamefont
  {Rosen}(2011)}]{Hassan:2011vm}%
  \BibitemOpen
  \bibfield  {author} {\bibinfo {author} {\bibfnamefont {S.}~\bibnamefont
  {Hassan}}\ and\ \bibinfo {author} {\bibfnamefont {R.~A.}\ \bibnamefont
  {Rosen}},\ }\href {\doibase 10.1007/JHEP07(2011)009} {\bibfield  {journal}
  {\bibinfo  {journal} {JHEP}\ }\textbf {\bibinfo {volume} {1107}},\ \bibinfo
  {pages} {009} (\bibinfo {year} {2011})},\ \Eprint
  {http://arxiv.org/abs/1103.6055} {arXiv:1103.6055 [hep-th]} \BibitemShut
  {NoStop}%
%%CITATION = ARXIV:1103.6055;%%
\bibitem [{\citenamefont {Hassan}\ and\ \citenamefont
  {Rosen}(2012{\natexlab{a}})}]{Hassan:2011hr}%
  \BibitemOpen
  \bibfield  {author} {\bibinfo {author} {\bibfnamefont {S.}~\bibnamefont
  {Hassan}}\ and\ \bibinfo {author} {\bibfnamefont {R.~A.}\ \bibnamefont
  {Rosen}},\ }\href {\doibase 10.1103/PhysRevLett.108.041101} {\bibfield
  {journal} {\bibinfo  {journal} {Phys.Rev.Lett.}\ }\textbf {\bibinfo {volume}
  {108}},\ \bibinfo {pages} {041101} (\bibinfo {year} {2012}{\natexlab{a}})},\
  \Eprint {http://arxiv.org/abs/1106.3344} {arXiv:1106.3344 [hep-th]}
  \BibitemShut {NoStop}%
%%CITATION = ARXIV:1106.3344;%%
\bibitem [{\citenamefont {Hassan}\ \emph {et~al.}(2012)\citenamefont {Hassan},
  \citenamefont {Rosen},\ and\ \citenamefont {Schmidt-May}}]{Hassan:2011tf}%
  \BibitemOpen
  \bibfield  {author} {\bibinfo {author} {\bibfnamefont {S.}~\bibnamefont
  {Hassan}}, \bibinfo {author} {\bibfnamefont {R.~A.}\ \bibnamefont {Rosen}}, \
  and\ \bibinfo {author} {\bibfnamefont {A.}~\bibnamefont {Schmidt-May}},\
  }\href {\doibase 10.1007/JHEP02(2012)026} {\bibfield  {journal} {\bibinfo
  {journal} {JHEP}\ }\textbf {\bibinfo {volume} {1202}},\ \bibinfo {pages}
  {026} (\bibinfo {year} {2012})},\ \Eprint {http://arxiv.org/abs/1109.3230}
  {arXiv:1109.3230 [hep-th]} \BibitemShut {NoStop}%
%%CITATION = ARXIV:1109.3230;%%
\bibitem [{\citenamefont {Hassan}\ and\ \citenamefont
  {Rosen}(2012{\natexlab{b}})}]{Hassan:2011zd}%
  \BibitemOpen
  \bibfield  {author} {\bibinfo {author} {\bibfnamefont {S.}~\bibnamefont
  {Hassan}}\ and\ \bibinfo {author} {\bibfnamefont {R.~A.}\ \bibnamefont
  {Rosen}},\ }\href {\doibase 10.1007/JHEP02(2012)126} {\bibfield  {journal}
  {\bibinfo  {journal} {JHEP}\ }\textbf {\bibinfo {volume} {1202}},\ \bibinfo
  {pages} {126} (\bibinfo {year} {2012}{\natexlab{b}})},\ \Eprint
  {http://arxiv.org/abs/1109.3515} {arXiv:1109.3515 [hep-th]} \BibitemShut
  {NoStop}%
%%CITATION = ARXIV:1109.3515;%%
\bibitem [{\citenamefont {Hassan}\ and\ \citenamefont
  {Rosen}(2012{\natexlab{c}})}]{Hassan:2011ea}%
  \BibitemOpen
  \bibfield  {author} {\bibinfo {author} {\bibfnamefont {S.}~\bibnamefont
  {Hassan}}\ and\ \bibinfo {author} {\bibfnamefont {R.~A.}\ \bibnamefont
  {Rosen}},\ }\href {\doibase 10.1007/JHEP04(2012)123} {\bibfield  {journal}
  {\bibinfo  {journal} {JHEP}\ }\textbf {\bibinfo {volume} {1204}},\ \bibinfo
  {pages} {123} (\bibinfo {year} {2012}{\natexlab{c}})},\ \Eprint
  {http://arxiv.org/abs/1111.2070} {arXiv:1111.2070 [hep-th]} \BibitemShut
  {NoStop}%
%%CITATION = ARXIV:1111.2070;%%
\bibitem [{\citenamefont {de~Rham}(2014)}]{deRham:2014zqa}%
  \BibitemOpen
  \bibfield  {author} {\bibinfo {author} {\bibfnamefont {C.}~\bibnamefont
  {de~Rham}},\ }\href {\doibase 10.12942/lrr-2014-7} {\bibfield  {journal}
  {\bibinfo  {journal} {Living Rev.Rel.}\ }\textbf {\bibinfo {volume} {17}},\
  \bibinfo {pages} {7} (\bibinfo {year} {2014})},\ \Eprint
  {http://arxiv.org/abs/1401.4173} {arXiv:1401.4173 [hep-th]} \BibitemShut
  {NoStop}%
%%CITATION = ARXIV:1401.4173;%%
\bibitem [{\citenamefont {D'Amico}\ \emph {et~al.}(2011)\citenamefont
  {D'Amico}, \citenamefont {de~Rham}, \citenamefont {Dubovsky}, \citenamefont
  {Gabadadze}, \citenamefont {Pirtskhalava} \emph {et~al.}}]{D'Amico:2011jj}%
  \BibitemOpen
  \bibfield  {author} {\bibinfo {author} {\bibfnamefont {G.}~\bibnamefont
  {D'Amico}}, \bibinfo {author} {\bibfnamefont {C.}~\bibnamefont {de~Rham}},
  \bibinfo {author} {\bibfnamefont {S.}~\bibnamefont {Dubovsky}}, \bibinfo
  {author} {\bibfnamefont {G.}~\bibnamefont {Gabadadze}}, \bibinfo {author}
  {\bibfnamefont {D.}~\bibnamefont {Pirtskhalava}},  \emph {et~al.},\ }\href
  {\doibase 10.1103/PhysRevD.84.124046} {\bibfield  {journal} {\bibinfo
  {journal} {Phys.Rev.}\ }\textbf {\bibinfo {volume} {D84}},\ \bibinfo {pages}
  {124046} (\bibinfo {year} {2011})},\ \Eprint {http://arxiv.org/abs/1108.5231}
  {arXiv:1108.5231 [hep-th]} \BibitemShut {NoStop}%
%%CITATION = ARXIV:1108.5231;%%
\bibitem [{\citenamefont {Higuchi}(1987)}]{Higuchi:1986py}%
  \BibitemOpen
  \bibfield  {author} {\bibinfo {author} {\bibfnamefont {A.}~\bibnamefont
  {Higuchi}},\ }\href {\doibase 10.1016/0550-3213(87)90691-2} {\bibfield
  {journal} {\bibinfo  {journal} {Nucl.Phys.}\ }\textbf {\bibinfo {volume}
  {B282}},\ \bibinfo {pages} {397} (\bibinfo {year} {1987})}\BibitemShut
  {NoStop}%
%%CITATION = NUPHA,B282,397;%%
\bibitem [{\citenamefont {Gumrukcuoglu}\ \emph {et~al.}(2011)\citenamefont
  {Gumrukcuoglu}, \citenamefont {Lin},\ and\ \citenamefont
  {Mukohyama}}]{Gumrukcuoglu:2011ew}%
  \BibitemOpen
  \bibfield  {author} {\bibinfo {author} {\bibfnamefont {A.~E.}\ \bibnamefont
  {Gumrukcuoglu}}, \bibinfo {author} {\bibfnamefont {C.}~\bibnamefont {Lin}}, \
  and\ \bibinfo {author} {\bibfnamefont {S.}~\bibnamefont {Mukohyama}},\ }\href
  {\doibase 10.1088/1475-7516/2011/11/030} {\bibfield  {journal} {\bibinfo
  {journal} {JCAP}\ }\textbf {\bibinfo {volume} {1111}},\ \bibinfo {pages}
  {030} (\bibinfo {year} {2011})},\ \Eprint {http://arxiv.org/abs/1109.3845}
  {arXiv:1109.3845 [hep-th]} \BibitemShut {NoStop}%
%%CITATION = ARXIV:1109.3845;%%
\bibitem [{\citenamefont {Gumrukcuoglu}\ \emph {et~al.}(2012)\citenamefont
  {Gumrukcuoglu}, \citenamefont {Lin},\ and\ \citenamefont
  {Mukohyama}}]{Gumrukcuoglu:2011zh}%
  \BibitemOpen
  \bibfield  {author} {\bibinfo {author} {\bibfnamefont {A.~E.}\ \bibnamefont
  {Gumrukcuoglu}}, \bibinfo {author} {\bibfnamefont {C.}~\bibnamefont {Lin}}, \
  and\ \bibinfo {author} {\bibfnamefont {S.}~\bibnamefont {Mukohyama}},\ }\href
  {\doibase 10.1088/1475-7516/2012/03/006} {\bibfield  {journal} {\bibinfo
  {journal} {JCAP}\ }\textbf {\bibinfo {volume} {1203}},\ \bibinfo {pages}
  {006} (\bibinfo {year} {2012})},\ \Eprint {http://arxiv.org/abs/1111.4107}
  {arXiv:1111.4107 [hep-th]} \BibitemShut {NoStop}%
%%CITATION = ARXIV:1111.4107;%%
\bibitem [{\citenamefont {Vakili}\ and\ \citenamefont
  {Khosravi}(2012)}]{Vakili:2012tm}%
  \BibitemOpen
  \bibfield  {author} {\bibinfo {author} {\bibfnamefont {B.}~\bibnamefont
  {Vakili}}\ and\ \bibinfo {author} {\bibfnamefont {N.}~\bibnamefont
  {Khosravi}},\ }\href {\doibase 10.1103/PhysRevD.85.083529} {\bibfield
  {journal} {\bibinfo  {journal} {Phys.Rev.}\ }\textbf {\bibinfo {volume}
  {D85}},\ \bibinfo {pages} {083529} (\bibinfo {year} {2012})},\ \Eprint
  {http://arxiv.org/abs/1204.1456} {arXiv:1204.1456 [gr-qc]} \BibitemShut
  {NoStop}%
%%CITATION = ARXIV:1204.1456;%%
\bibitem [{\citenamefont {De~Felice}\ \emph {et~al.}(2012)\citenamefont
  {De~Felice}, \citenamefont {G{\"u}mr{\"u}k{\c c}{\"u}o{\u g}lu},\ and\
  \citenamefont {Mukohyama}}]{DeFelice:2012mx}%
  \BibitemOpen
  \bibfield  {author} {\bibinfo {author} {\bibfnamefont {A.}~\bibnamefont
  {De~Felice}}, \bibinfo {author} {\bibfnamefont {A.~E.}\ \bibnamefont
  {G{\"u}mr{\"u}k{\c c}{\"u}o{\u g}lu}}, \ and\ \bibinfo {author}
  {\bibfnamefont {S.}~\bibnamefont {Mukohyama}},\ }\href {\doibase
  10.1103/PhysRevLett.109.171101} {\bibfield  {journal} {\bibinfo  {journal}
  {Phys.Rev.Lett.}\ }\textbf {\bibinfo {volume} {109}},\ \bibinfo {pages}
  {171101} (\bibinfo {year} {2012})},\ \Eprint {http://arxiv.org/abs/1206.2080}
  {arXiv:1206.2080 [hep-th]} \BibitemShut {NoStop}%
%%CITATION = ARXIV:1206.2080;%%
\bibitem [{\citenamefont {Fasiello}\ and\ \citenamefont
  {Tolley}(2012)}]{Fasiello:2012rw}%
  \BibitemOpen
  \bibfield  {author} {\bibinfo {author} {\bibfnamefont {M.}~\bibnamefont
  {Fasiello}}\ and\ \bibinfo {author} {\bibfnamefont {A.~J.}\ \bibnamefont
  {Tolley}},\ }\href {\doibase 10.1088/1475-7516/2012/11/035} {\bibfield
  {journal} {\bibinfo  {journal} {JCAP}\ }\textbf {\bibinfo {volume} {1211}},\
  \bibinfo {pages} {035} (\bibinfo {year} {2012})},\ \Eprint
  {http://arxiv.org/abs/1206.3852} {arXiv:1206.3852 [hep-th]} \BibitemShut
  {NoStop}%
%%CITATION = ARXIV:1206.3852;%%
\bibitem [{\citenamefont {De~Felice}\ \emph
  {et~al.}(2013{\natexlab{a}})\citenamefont {De~Felice}, \citenamefont
  {G{\"u}mr{\"u}k{\c c}{\"u}o{\u g}lu}, \citenamefont {Lin},\ and\
  \citenamefont {Mukohyama}}]{DeFelice:2013awa}%
  \BibitemOpen
  \bibfield  {author} {\bibinfo {author} {\bibfnamefont {A.}~\bibnamefont
  {De~Felice}}, \bibinfo {author} {\bibfnamefont {A.~E.}\ \bibnamefont
  {G{\"u}mr{\"u}k{\c c}{\"u}o{\u g}lu}}, \bibinfo {author} {\bibfnamefont
  {C.}~\bibnamefont {Lin}}, \ and\ \bibinfo {author} {\bibfnamefont
  {S.}~\bibnamefont {Mukohyama}},\ }\href {\doibase
  10.1088/1475-7516/2013/05/035} {\bibfield  {journal} {\bibinfo  {journal}
  {JCAP}\ }\textbf {\bibinfo {volume} {1305}},\ \bibinfo {pages} {035}
  (\bibinfo {year} {2013}{\natexlab{a}})},\ \Eprint
  {http://arxiv.org/abs/1303.4154} {arXiv:1303.4154 [hep-th]} \BibitemShut
  {NoStop}%
%%CITATION = ARXIV:1303.4154;%%
\bibitem [{\citenamefont {Gratia}\ \emph {et~al.}(2012)\citenamefont {Gratia},
  \citenamefont {Hu},\ and\ \citenamefont {Wyman}}]{Gratia:2012wt}%
  \BibitemOpen
  \bibfield  {author} {\bibinfo {author} {\bibfnamefont {P.}~\bibnamefont
  {Gratia}}, \bibinfo {author} {\bibfnamefont {W.}~\bibnamefont {Hu}}, \ and\
  \bibinfo {author} {\bibfnamefont {M.}~\bibnamefont {Wyman}},\ }\href
  {\doibase 10.1103/PhysRevD.86.061504} {\bibfield  {journal} {\bibinfo
  {journal} {Phys.Rev.}\ }\textbf {\bibinfo {volume} {D86}},\ \bibinfo {pages}
  {061504} (\bibinfo {year} {2012})},\ \Eprint {http://arxiv.org/abs/1205.4241}
  {arXiv:1205.4241 [hep-th]} \BibitemShut {NoStop}%
%%CITATION = ARXIV:1205.4241;%%
\bibitem [{\citenamefont {G{\"u}mr{\"u}k{\c c}{\"u}o{\u g}lu}\ \emph
  {et~al.}(2012)\citenamefont {G{\"u}mr{\"u}k{\c c}{\"u}o{\u g}lu},
  \citenamefont {Lin},\ and\ \citenamefont {Mukohyama}}]{Gumrukcuoglu:2012aa}%
  \BibitemOpen
  \bibfield  {author} {\bibinfo {author} {\bibfnamefont {A.~E.}\ \bibnamefont
  {G{\"u}mr{\"u}k{\c c}{\"u}o{\u g}lu}}, \bibinfo {author} {\bibfnamefont
  {C.}~\bibnamefont {Lin}}, \ and\ \bibinfo {author} {\bibfnamefont
  {S.}~\bibnamefont {Mukohyama}},\ }\href {\doibase
  10.1016/j.physletb.2012.09.049} {\bibfield  {journal} {\bibinfo  {journal}
  {Phys.Lett.}\ }\textbf {\bibinfo {volume} {B717}},\ \bibinfo {pages} {295}
  (\bibinfo {year} {2012})},\ \Eprint {http://arxiv.org/abs/1206.2723}
  {arXiv:1206.2723 [hep-th]} \BibitemShut {NoStop}%
%%CITATION = ARXIV:1206.2723;%%
\bibitem [{\citenamefont {Volkov}(2012{\natexlab{a}})}]{Volkov:2012cf}%
  \BibitemOpen
  \bibfield  {author} {\bibinfo {author} {\bibfnamefont {M.~S.}\ \bibnamefont
  {Volkov}},\ }\href {\doibase 10.1103/PhysRevD.86.061502} {\bibfield
  {journal} {\bibinfo  {journal} {Phys.Rev.}\ }\textbf {\bibinfo {volume}
  {D86}},\ \bibinfo {pages} {061502} (\bibinfo {year} {2012}{\natexlab{a}})},\
  \Eprint {http://arxiv.org/abs/1205.5713} {arXiv:1205.5713 [hep-th]}
  \BibitemShut {NoStop}%
%%CITATION = ARXIV:1205.5713;%%
\bibitem [{\citenamefont {Volkov}(2012{\natexlab{b}})}]{Volkov:2012zb}%
  \BibitemOpen
  \bibfield  {author} {\bibinfo {author} {\bibfnamefont {M.~S.}\ \bibnamefont
  {Volkov}},\ }\href {\doibase 10.1103/PhysRevD.86.104022} {\bibfield
  {journal} {\bibinfo  {journal} {Phys.Rev.}\ }\textbf {\bibinfo {volume}
  {D86}},\ \bibinfo {pages} {104022} (\bibinfo {year} {2012}{\natexlab{b}})},\
  \Eprint {http://arxiv.org/abs/1207.3723} {arXiv:1207.3723 [hep-th]}
  \BibitemShut {NoStop}%
%%CITATION = ARXIV:1207.3723;%%
\bibitem [{\citenamefont {De~Felice}\ \emph
  {et~al.}(2013{\natexlab{b}})\citenamefont {De~Felice}, \citenamefont
  {G{\"u}mr{\"u}k{\c c}{\"u}o{\u g}lu}, \citenamefont {Lin},\ and\
  \citenamefont {Mukohyama}}]{DeFelice:2013bxa}%
  \BibitemOpen
  \bibfield  {author} {\bibinfo {author} {\bibfnamefont {A.}~\bibnamefont
  {De~Felice}}, \bibinfo {author} {\bibfnamefont {A.~E.}\ \bibnamefont
  {G{\"u}mr{\"u}k{\c c}{\"u}o{\u g}lu}}, \bibinfo {author} {\bibfnamefont
  {C.}~\bibnamefont {Lin}}, \ and\ \bibinfo {author} {\bibfnamefont
  {S.}~\bibnamefont {Mukohyama}},\ }\href {\doibase
  10.1088/0264-9381/30/18/184004} {\bibfield  {journal} {\bibinfo  {journal}
  {Class.Quant.Grav.}\ }\textbf {\bibinfo {volume} {30}},\ \bibinfo {pages}
  {184004} (\bibinfo {year} {2013}{\natexlab{b}})},\ \Eprint
  {http://arxiv.org/abs/1304.0484} {arXiv:1304.0484 [hep-th]} \BibitemShut
  {NoStop}%
%%CITATION = ARXIV:1304.0484;%%
\bibitem [{\citenamefont {Do}\ and\ \citenamefont {Kao}(2013)}]{Do:2013tea}%
  \BibitemOpen
  \bibfield  {author} {\bibinfo {author} {\bibfnamefont {T.~Q.}\ \bibnamefont
  {Do}}\ and\ \bibinfo {author} {\bibfnamefont {W.}~\bibnamefont {Kao}},\
  }\href {\doibase 10.1103/PhysRevD.88.063006} {\bibfield  {journal} {\bibinfo
  {journal} {Phys.Rev.}\ }\textbf {\bibinfo {volume} {D88}},\ \bibinfo {pages}
  {063006} (\bibinfo {year} {2013})}\BibitemShut {NoStop}%
%%CITATION = PHRVA,D88,063006;%%
\bibitem [{\citenamefont {Kao}\ and\ \citenamefont {Lin}(2014)}]{Kao:2014efa}%
  \BibitemOpen
  \bibfield  {author} {\bibinfo {author} {\bibfnamefont {W.~â.}\ \bibnamefont
  {Kao}}\ and\ \bibinfo {author} {\bibfnamefont {I.~C.}\ \bibnamefont {Lin}},\
  }\href {\doibase 10.1103/PhysRevD.90.063003} {\bibfield  {journal} {\bibinfo
  {journal} {Phys.Rev.}\ }\textbf {\bibinfo {volume} {D90}},\ \bibinfo {pages}
  {063003} (\bibinfo {year} {2014})}\BibitemShut {NoStop}%
%%CITATION = PHRVA,D90,063003;%%
\bibitem [{\citenamefont {de~Rham}\ \emph
  {et~al.}(2014{\natexlab{a}})\citenamefont {de~Rham}, \citenamefont
  {Fasiello},\ and\ \citenamefont {Tolley}}]{deRham:2014gla}%
  \BibitemOpen
  \bibfield  {author} {\bibinfo {author} {\bibfnamefont {C.}~\bibnamefont
  {de~Rham}}, \bibinfo {author} {\bibfnamefont {M.}~\bibnamefont {Fasiello}}, \
  and\ \bibinfo {author} {\bibfnamefont {A.~J.}\ \bibnamefont {Tolley}},\
  }\href@noop {} {\  (\bibinfo {year} {2014}{\natexlab{a}})},\ \Eprint
  {http://arxiv.org/abs/1410.0960} {arXiv:1410.0960 [hep-th]} \BibitemShut
  {NoStop}%
%%CITATION = ARXIV:1410.0960;%%
\bibitem [{\citenamefont {D'Amico}\ \emph {et~al.}(2013)\citenamefont
  {D'Amico}, \citenamefont {Gabadadze}, \citenamefont {Hui},\ and\
  \citenamefont {Pirtskhalava}}]{D'Amico:2012zv}%
  \BibitemOpen
  \bibfield  {author} {\bibinfo {author} {\bibfnamefont {G.}~\bibnamefont
  {D'Amico}}, \bibinfo {author} {\bibfnamefont {G.}~\bibnamefont {Gabadadze}},
  \bibinfo {author} {\bibfnamefont {L.}~\bibnamefont {Hui}}, \ and\ \bibinfo
  {author} {\bibfnamefont {D.}~\bibnamefont {Pirtskhalava}},\ }\href {\doibase
  10.1103/PhysRevD.87.064037} {\bibfield  {journal} {\bibinfo  {journal}
  {Phys.Rev.}\ }\textbf {\bibinfo {volume} {D87}},\ \bibinfo {pages} {064037}
  (\bibinfo {year} {2013})},\ \Eprint {http://arxiv.org/abs/1206.4253}
  {arXiv:1206.4253 [hep-th]} \BibitemShut {NoStop}%
%%CITATION = ARXIV:1206.4253;%%
\bibitem [{\citenamefont {Huang}\ \emph {et~al.}(2012)\citenamefont {Huang},
  \citenamefont {Piao},\ and\ \citenamefont {Zhou}}]{Huang:2012pe}%
  \BibitemOpen
  \bibfield  {author} {\bibinfo {author} {\bibfnamefont {Q.-G.}\ \bibnamefont
  {Huang}}, \bibinfo {author} {\bibfnamefont {Y.-S.}\ \bibnamefont {Piao}}, \
  and\ \bibinfo {author} {\bibfnamefont {S.-Y.}\ \bibnamefont {Zhou}},\ }\href
  {\doibase 10.1103/PhysRevD.86.124014} {\bibfield  {journal} {\bibinfo
  {journal} {Phys.Rev.}\ }\textbf {\bibinfo {volume} {D86}},\ \bibinfo {pages}
  {124014} (\bibinfo {year} {2012})},\ \Eprint {http://arxiv.org/abs/1206.5678}
  {arXiv:1206.5678 [hep-th]} \BibitemShut {NoStop}%
%%CITATION = ARXIV:1206.5678;%%
\bibitem [{\citenamefont {Jaccard}\ \emph {et~al.}(2013)\citenamefont
  {Jaccard}, \citenamefont {Maggiore},\ and\ \citenamefont
  {Mitsou}}]{Jaccard:2013gla}%
  \BibitemOpen
  \bibfield  {author} {\bibinfo {author} {\bibfnamefont {M.}~\bibnamefont
  {Jaccard}}, \bibinfo {author} {\bibfnamefont {M.}~\bibnamefont {Maggiore}}, \
  and\ \bibinfo {author} {\bibfnamefont {E.}~\bibnamefont {Mitsou}},\ }\href
  {\doibase 10.1103/PhysRevD.88.044033} {\bibfield  {journal} {\bibinfo
  {journal} {Phys.Rev.}\ }\textbf {\bibinfo {volume} {D88}},\ \bibinfo {pages}
  {044033} (\bibinfo {year} {2013})},\ \Eprint {http://arxiv.org/abs/1305.3034}
  {arXiv:1305.3034 [hep-th]} \BibitemShut {NoStop}%
%%CITATION = ARXIV:1305.3034;%%
\bibitem [{\citenamefont {Foffa}\ \emph {et~al.}(2014)\citenamefont {Foffa},
  \citenamefont {Maggiore},\ and\ \citenamefont {Mitsou}}]{Foffa:2013vma}%
  \BibitemOpen
  \bibfield  {author} {\bibinfo {author} {\bibfnamefont {S.}~\bibnamefont
  {Foffa}}, \bibinfo {author} {\bibfnamefont {M.}~\bibnamefont {Maggiore}}, \
  and\ \bibinfo {author} {\bibfnamefont {E.}~\bibnamefont {Mitsou}},\ }\href
  {\doibase 10.1142/S0217751X14501164} {\bibfield  {journal} {\bibinfo
  {journal} {Int.J.Mod.Phys.}\ }\textbf {\bibinfo {volume} {A29}},\ \bibinfo
  {pages} {1450116} (\bibinfo {year} {2014})},\ \Eprint
  {http://arxiv.org/abs/1311.3435} {arXiv:1311.3435 [hep-th]} \BibitemShut
  {NoStop}%
%%CITATION = ARXIV:1311.3435;%%
\bibitem [{\citenamefont {Dirian}\ \emph {et~al.}(2014)\citenamefont {Dirian},
  \citenamefont {Foffa}, \citenamefont {Khosravi}, \citenamefont {Kunz},\ and\
  \citenamefont {Maggiore}}]{Dirian:2014ara}%
  \BibitemOpen
  \bibfield  {author} {\bibinfo {author} {\bibfnamefont {Y.}~\bibnamefont
  {Dirian}}, \bibinfo {author} {\bibfnamefont {S.}~\bibnamefont {Foffa}},
  \bibinfo {author} {\bibfnamefont {N.}~\bibnamefont {Khosravi}}, \bibinfo
  {author} {\bibfnamefont {M.}~\bibnamefont {Kunz}}, \ and\ \bibinfo {author}
  {\bibfnamefont {M.}~\bibnamefont {Maggiore}},\ }\href {\doibase
  10.1088/1475-7516/2014/06/033} {\bibfield  {journal} {\bibinfo  {journal}
  {JCAP}\ }\textbf {\bibinfo {volume} {1406}},\ \bibinfo {pages} {033}
  (\bibinfo {year} {2014})},\ \Eprint {http://arxiv.org/abs/1403.6068}
  {arXiv:1403.6068 [astro-ph.CO]} \BibitemShut {NoStop}%
%%CITATION = ARXIV:1403.6068;%%
\bibitem [{\citenamefont {Conroy}\ \emph {et~al.}(2015)\citenamefont {Conroy},
  \citenamefont {Koivisto}, \citenamefont {Mazumdar},\ and\ \citenamefont
  {Teimouri}}]{Conroy:2014eja}%
  \BibitemOpen
  \bibfield  {author} {\bibinfo {author} {\bibfnamefont {A.}~\bibnamefont
  {Conroy}}, \bibinfo {author} {\bibfnamefont {T.}~\bibnamefont {Koivisto}},
  \bibinfo {author} {\bibfnamefont {A.}~\bibnamefont {Mazumdar}}, \ and\
  \bibinfo {author} {\bibfnamefont {A.}~\bibnamefont {Teimouri}},\ }\href
  {\doibase 10.1088/0264-9381/32/1/015024} {\bibfield  {journal} {\bibinfo
  {journal} {Class.Quant.Grav.}\ }\textbf {\bibinfo {volume} {32}},\ \bibinfo
  {pages} {015024} (\bibinfo {year} {2015})},\ \Eprint
  {http://arxiv.org/abs/1406.4998} {arXiv:1406.4998 [hep-th]} \BibitemShut
  {NoStop}%
%%CITATION = ARXIV:1406.4998;%%
\bibitem [{\citenamefont {Comelli}\ \emph {et~al.}(2013)\citenamefont
  {Comelli}, \citenamefont {Nesti},\ and\ \citenamefont
  {Pilo}}]{Comelli:2013paa}%
  \BibitemOpen
  \bibfield  {author} {\bibinfo {author} {\bibfnamefont {D.}~\bibnamefont
  {Comelli}}, \bibinfo {author} {\bibfnamefont {F.}~\bibnamefont {Nesti}}, \
  and\ \bibinfo {author} {\bibfnamefont {L.}~\bibnamefont {Pilo}},\ }\href
  {\doibase 10.1103/PhysRevD.87.124021} {\bibfield  {journal} {\bibinfo
  {journal} {Phys.Rev.}\ }\textbf {\bibinfo {volume} {D87}},\ \bibinfo {pages}
  {124021} (\bibinfo {year} {2013})},\ \Eprint {http://arxiv.org/abs/1302.4447}
  {arXiv:1302.4447 [hep-th]} \BibitemShut {NoStop}%
%%CITATION = ARXIV:1302.4447;%%
\bibitem [{\citenamefont {Comelli}\ \emph
  {et~al.}(2014{\natexlab{a}})\citenamefont {Comelli}, \citenamefont {Nesti},\
  and\ \citenamefont {Pilo}}]{Comelli:2013tja}%
  \BibitemOpen
  \bibfield  {author} {\bibinfo {author} {\bibfnamefont {D.}~\bibnamefont
  {Comelli}}, \bibinfo {author} {\bibfnamefont {F.}~\bibnamefont {Nesti}}, \
  and\ \bibinfo {author} {\bibfnamefont {L.}~\bibnamefont {Pilo}},\ }\href
  {\doibase 10.1088/1475-7516/2014/05/036} {\bibfield  {journal} {\bibinfo
  {journal} {JCAP}\ }\textbf {\bibinfo {volume} {1405}},\ \bibinfo {pages}
  {036} (\bibinfo {year} {2014}{\natexlab{a}})},\ \Eprint
  {http://arxiv.org/abs/1307.8329} {arXiv:1307.8329 [hep-th]} \BibitemShut
  {NoStop}%
%%CITATION = ARXIV:1307.8329;%%
\bibitem [{\citenamefont {de~Rham}\ \emph
  {et~al.}(2014{\natexlab{b}})\citenamefont {de~Rham}, \citenamefont
  {Heisenberg},\ and\ \citenamefont {Ribeiro}}]{deRham:2014naa}%
  \BibitemOpen
  \bibfield  {author} {\bibinfo {author} {\bibfnamefont {C.}~\bibnamefont
  {de~Rham}}, \bibinfo {author} {\bibfnamefont {L.}~\bibnamefont {Heisenberg}},
  \ and\ \bibinfo {author} {\bibfnamefont {R.~H.}\ \bibnamefont {Ribeiro}},\
  }\href@noop {} {\  (\bibinfo {year} {2014}{\natexlab{b}})},\ \Eprint
  {http://arxiv.org/abs/1408.1678} {arXiv:1408.1678 [hep-th]} \BibitemShut
  {NoStop}%
%%CITATION = ARXIV:1408.1678;%%
\bibitem [{\citenamefont {de~Rham}\ \emph
  {et~al.}(2014{\natexlab{c}})\citenamefont {de~Rham}, \citenamefont
  {Heisenberg},\ and\ \citenamefont {Ribeiro}}]{deRham:2014fha}%
  \BibitemOpen
  \bibfield  {author} {\bibinfo {author} {\bibfnamefont {C.}~\bibnamefont
  {de~Rham}}, \bibinfo {author} {\bibfnamefont {L.}~\bibnamefont {Heisenberg}},
  \ and\ \bibinfo {author} {\bibfnamefont {R.~H.}\ \bibnamefont {Ribeiro}},\
  }\href@noop {} {\  (\bibinfo {year} {2014}{\natexlab{c}})},\ \Eprint
  {http://arxiv.org/abs/1409.3834} {arXiv:1409.3834 [hep-th]} \BibitemShut
  {NoStop}%
%%CITATION = ARXIV:1409.3834;%%
\bibitem [{\citenamefont {G{\"u}mr{\"u}k{\c c}{\"u}o{\u g}lu}\ \emph
  {et~al.}(2014)\citenamefont {G{\"u}mr{\"u}k{\c c}{\"u}o{\u g}lu},
  \citenamefont {Heisenberg},\ and\ \citenamefont
  {Mukohyama}}]{Gumrukcuoglu:2014xba}%
  \BibitemOpen
  \bibfield  {author} {\bibinfo {author} {\bibfnamefont {A.~E.}\ \bibnamefont
  {G{\"u}mr{\"u}k{\c c}{\"u}o{\u g}lu}}, \bibinfo {author} {\bibfnamefont
  {L.}~\bibnamefont {Heisenberg}}, \ and\ \bibinfo {author} {\bibfnamefont
  {S.}~\bibnamefont {Mukohyama}},\ }\href@noop {} {\  (\bibinfo {year}
  {2014})},\ \Eprint {http://arxiv.org/abs/1409.7260} {arXiv:1409.7260
  [hep-th]} \BibitemShut {NoStop}%
%%CITATION = ARXIV:1409.7260;%%
\bibitem [{\citenamefont {Solomon}\ \emph
  {et~al.}(2014{\natexlab{a}})\citenamefont {Solomon}, \citenamefont {Enander},
  \citenamefont {Akrami}, \citenamefont {Koivisto}, \citenamefont {K{\"o}nnig}
  \emph {et~al.}}]{Solomon:2014iwa}%
  \BibitemOpen
  \bibfield  {author} {\bibinfo {author} {\bibfnamefont {A.~R.}\ \bibnamefont
  {Solomon}}, \bibinfo {author} {\bibfnamefont {J.}~\bibnamefont {Enander}},
  \bibinfo {author} {\bibfnamefont {Y.}~\bibnamefont {Akrami}}, \bibinfo
  {author} {\bibfnamefont {T.~S.}\ \bibnamefont {Koivisto}}, \bibinfo {author}
  {\bibfnamefont {F.}~\bibnamefont {K{\"o}nnig}},  \emph {et~al.},\ }\href@noop
  {} {\  (\bibinfo {year} {2014}{\natexlab{a}})},\ \Eprint
  {http://arxiv.org/abs/1409.8300} {arXiv:1409.8300 [astro-ph.CO]} \BibitemShut
  {NoStop}%
%%CITATION = ARXIV:1409.8300;%%
\bibitem [{\citenamefont {Yamashita}\ \emph {et~al.}(2014)\citenamefont
  {Yamashita}, \citenamefont {De~Felice},\ and\ \citenamefont
  {Tanaka}}]{Yamashita:2014fga}%
  \BibitemOpen
  \bibfield  {author} {\bibinfo {author} {\bibfnamefont {Y.}~\bibnamefont
  {Yamashita}}, \bibinfo {author} {\bibfnamefont {A.}~\bibnamefont
  {De~Felice}}, \ and\ \bibinfo {author} {\bibfnamefont {T.}~\bibnamefont
  {Tanaka}},\ }\href@noop {} {\  (\bibinfo {year} {2014})},\ \Eprint
  {http://arxiv.org/abs/1408.0487} {arXiv:1408.0487 [hep-th]} \BibitemShut
  {NoStop}%
%%CITATION = ARXIV:1408.0487;%%
\bibitem [{\citenamefont {Noller}\ and\ \citenamefont
  {Melville}(2014)}]{Noller:2014sta}%
  \BibitemOpen
  \bibfield  {author} {\bibinfo {author} {\bibfnamefont {J.}~\bibnamefont
  {Noller}}\ and\ \bibinfo {author} {\bibfnamefont {S.}~\bibnamefont
  {Melville}},\ }\href@noop {} {\  (\bibinfo {year} {2014})},\ \Eprint
  {http://arxiv.org/abs/1408.5131} {arXiv:1408.5131 [hep-th]} \BibitemShut
  {NoStop}%
%%CITATION = ARXIV:1408.5131;%%
\bibitem [{\citenamefont {Hassan}\ \emph {et~al.}(2014)\citenamefont {Hassan},
  \citenamefont {Kocic},\ and\ \citenamefont {Schmidt-May}}]{Hassan:2014gta}%
  \BibitemOpen
  \bibfield  {author} {\bibinfo {author} {\bibfnamefont {S.}~\bibnamefont
  {Hassan}}, \bibinfo {author} {\bibfnamefont {M.}~\bibnamefont {Kocic}}, \
  and\ \bibinfo {author} {\bibfnamefont {A.}~\bibnamefont {Schmidt-May}},\
  }\href@noop {} {\  (\bibinfo {year} {2014})},\ \Eprint
  {http://arxiv.org/abs/1409.1909} {arXiv:1409.1909 [hep-th]} \BibitemShut
  {NoStop}%
%%CITATION = ARXIV:1409.1909;%%
\bibitem [{\citenamefont {Soloviev}(2014)}]{Soloviev:2014eea}%
  \BibitemOpen
  \bibfield  {author} {\bibinfo {author} {\bibfnamefont {V.~O.}\ \bibnamefont
  {Soloviev}},\ }\href@noop {} {\  (\bibinfo {year} {2014})},\ \Eprint
  {http://arxiv.org/abs/1410.0048} {arXiv:1410.0048 [hep-th]} \BibitemShut
  {NoStop}%
%%CITATION = ARXIV:1410.0048;%%
\bibitem [{\citenamefont {Heisenberg}(2014)}]{Heisenberg:2014rka}%
  \BibitemOpen
  \bibfield  {author} {\bibinfo {author} {\bibfnamefont {L.}~\bibnamefont
  {Heisenberg}},\ }\href@noop {} {\  (\bibinfo {year} {2014})},\ \Eprint
  {http://arxiv.org/abs/1410.4239} {arXiv:1410.4239 [hep-th]} \BibitemShut
  {NoStop}%
%%CITATION = ARXIV:1410.4239;%%
\bibitem [{\citenamefont {Akrami}\ \emph
  {et~al.}(2013{\natexlab{a}})\citenamefont {Akrami}, \citenamefont {Koivisto},
  \citenamefont {Mota},\ and\ \citenamefont {Sandstad}}]{Akrami:2013ffa}%
  \BibitemOpen
  \bibfield  {author} {\bibinfo {author} {\bibfnamefont {Y.}~\bibnamefont
  {Akrami}}, \bibinfo {author} {\bibfnamefont {T.~S.}\ \bibnamefont
  {Koivisto}}, \bibinfo {author} {\bibfnamefont {D.~F.}\ \bibnamefont {Mota}},
  \ and\ \bibinfo {author} {\bibfnamefont {M.}~\bibnamefont {Sandstad}},\
  }\href {\doibase 10.1088/1475-7516/2013/10/046} {\bibfield  {journal}
  {\bibinfo  {journal} {JCAP}\ }\textbf {\bibinfo {volume} {1310}},\ \bibinfo
  {pages} {046} (\bibinfo {year} {2013}{\natexlab{a}})},\ \Eprint
  {http://arxiv.org/abs/1306.0004} {arXiv:1306.0004 [hep-th]} \BibitemShut
  {NoStop}%
%%CITATION = ARXIV:1306.0004;%%
\bibitem [{\citenamefont {Akrami}\ \emph {et~al.}(2014)\citenamefont {Akrami},
  \citenamefont {Koivisto},\ and\ \citenamefont {Solomon}}]{Akrami:2014lja}%
  \BibitemOpen
  \bibfield  {author} {\bibinfo {author} {\bibfnamefont {Y.}~\bibnamefont
  {Akrami}}, \bibinfo {author} {\bibfnamefont {T.~S.}\ \bibnamefont
  {Koivisto}}, \ and\ \bibinfo {author} {\bibfnamefont {A.~R.}\ \bibnamefont
  {Solomon}},\ }\href {\doibase 10.1007/s10714-014-1838-4} {\bibfield
  {journal} {\bibinfo  {journal} {Gen.Rel.Grav.}\ }\textbf {\bibinfo {volume}
  {47}},\ \bibinfo {pages} {1838} (\bibinfo {year} {2014})},\ \Eprint
  {http://arxiv.org/abs/1404.0006} {arXiv:1404.0006 [gr-qc]} \BibitemShut
  {NoStop}%
%%CITATION = ARXIV:1404.0006;%%
\bibitem [{\citenamefont {Khosravi}\ \emph
  {et~al.}(2012{\natexlab{a}})\citenamefont {Khosravi}, \citenamefont
  {Rahmanpour}, \citenamefont {Sepangi},\ and\ \citenamefont
  {Shahidi}}]{Khosravi:2011zi}%
  \BibitemOpen
  \bibfield  {author} {\bibinfo {author} {\bibfnamefont {N.}~\bibnamefont
  {Khosravi}}, \bibinfo {author} {\bibfnamefont {N.}~\bibnamefont
  {Rahmanpour}}, \bibinfo {author} {\bibfnamefont {H.~R.}\ \bibnamefont
  {Sepangi}}, \ and\ \bibinfo {author} {\bibfnamefont {S.}~\bibnamefont
  {Shahidi}},\ }\href {\doibase 10.1103/PhysRevD.85.024049} {\bibfield
  {journal} {\bibinfo  {journal} {Phys.Rev.}\ }\textbf {\bibinfo {volume}
  {D85}},\ \bibinfo {pages} {024049} (\bibinfo {year} {2012}{\natexlab{a}})},\
  \Eprint {http://arxiv.org/abs/1111.5346} {arXiv:1111.5346 [hep-th]}
  \BibitemShut {NoStop}%
%%CITATION = ARXIV:1111.5346;%%
\bibitem [{\citenamefont {Tamanini}\ \emph {et~al.}(2014)\citenamefont
  {Tamanini}, \citenamefont {Saridakis},\ and\ \citenamefont
  {Koivisto}}]{Tamanini:2013xia}%
  \BibitemOpen
  \bibfield  {author} {\bibinfo {author} {\bibfnamefont {N.}~\bibnamefont
  {Tamanini}}, \bibinfo {author} {\bibfnamefont {E.~N.}\ \bibnamefont
  {Saridakis}}, \ and\ \bibinfo {author} {\bibfnamefont {T.~S.}\ \bibnamefont
  {Koivisto}},\ }\href {\doibase 10.1088/1475-7516/2014/02/015} {\bibfield
  {journal} {\bibinfo  {journal} {JCAP}\ }\textbf {\bibinfo {volume} {1402}},\
  \bibinfo {pages} {015} (\bibinfo {year} {2014})},\ \Eprint
  {http://arxiv.org/abs/1307.5984} {arXiv:1307.5984 [hep-th]} \BibitemShut
  {NoStop}%
%%CITATION = ARXIV:1307.5984;%%
\bibitem [{\citenamefont {Volkov}(2012{\natexlab{c}})}]{Volkov:2011an}%
  \BibitemOpen
  \bibfield  {author} {\bibinfo {author} {\bibfnamefont {M.~S.}\ \bibnamefont
  {Volkov}},\ }\href {\doibase 10.1007/JHEP01(2012)035} {\bibfield  {journal}
  {\bibinfo  {journal} {JHEP}\ }\textbf {\bibinfo {volume} {1201}},\ \bibinfo
  {pages} {035} (\bibinfo {year} {2012}{\natexlab{c}})},\ \Eprint
  {http://arxiv.org/abs/1110.6153} {arXiv:1110.6153 [hep-th]} \BibitemShut
  {NoStop}%
%%CITATION = ARXIV:1110.6153;%%
\bibitem [{\citenamefont {von Strauss}\ \emph {et~al.}(2012)\citenamefont {von
  Strauss}, \citenamefont {Schmidt-May}, \citenamefont {Enander}, \citenamefont
  {M{\"o}rtsell},\ and\ \citenamefont {Hassan}}]{vonStrauss:2011mq}%
  \BibitemOpen
  \bibfield  {author} {\bibinfo {author} {\bibfnamefont {M.}~\bibnamefont {von
  Strauss}}, \bibinfo {author} {\bibfnamefont {A.}~\bibnamefont {Schmidt-May}},
  \bibinfo {author} {\bibfnamefont {J.}~\bibnamefont {Enander}}, \bibinfo
  {author} {\bibfnamefont {E.}~\bibnamefont {M{\"o}rtsell}}, \ and\ \bibinfo
  {author} {\bibfnamefont {S.}~\bibnamefont {Hassan}},\ }\href {\doibase
  10.1088/1475-7516/2012/03/042} {\bibfield  {journal} {\bibinfo  {journal}
  {JCAP}\ }\textbf {\bibinfo {volume} {1203}},\ \bibinfo {pages} {042}
  (\bibinfo {year} {2012})},\ \Eprint {http://arxiv.org/abs/1111.1655}
  {arXiv:1111.1655 [gr-qc]} \BibitemShut {NoStop}%
%%CITATION = ARXIV:1111.1655;%%
\bibitem [{\citenamefont {Comelli}\ \emph
  {et~al.}(2012{\natexlab{a}})\citenamefont {Comelli}, \citenamefont
  {Crisostomi}, \citenamefont {Nesti},\ and\ \citenamefont
  {Pilo}}]{Comelli:2011zm}%
  \BibitemOpen
  \bibfield  {author} {\bibinfo {author} {\bibfnamefont {D.}~\bibnamefont
  {Comelli}}, \bibinfo {author} {\bibfnamefont {M.}~\bibnamefont {Crisostomi}},
  \bibinfo {author} {\bibfnamefont {F.}~\bibnamefont {Nesti}}, \ and\ \bibinfo
  {author} {\bibfnamefont {L.}~\bibnamefont {Pilo}},\ }\href {\doibase
  10.1007/JHEP06(2012)020, 10.1007/JHEP03(2012)067} {\bibfield  {journal}
  {\bibinfo  {journal} {JHEP}\ }\textbf {\bibinfo {volume} {1203}},\ \bibinfo
  {pages} {067} (\bibinfo {year} {2012}{\natexlab{a}})},\ \Eprint
  {http://arxiv.org/abs/1111.1983} {arXiv:1111.1983 [hep-th]} \BibitemShut
  {NoStop}%
%%CITATION = ARXIV:1111.1983;%%
\bibitem [{\citenamefont {Comelli}\ \emph
  {et~al.}(2012{\natexlab{b}})\citenamefont {Comelli}, \citenamefont
  {Crisostomi},\ and\ \citenamefont {Pilo}}]{Comelli:2012db}%
  \BibitemOpen
  \bibfield  {author} {\bibinfo {author} {\bibfnamefont {D.}~\bibnamefont
  {Comelli}}, \bibinfo {author} {\bibfnamefont {M.}~\bibnamefont {Crisostomi}},
  \ and\ \bibinfo {author} {\bibfnamefont {L.}~\bibnamefont {Pilo}},\ }\href
  {\doibase 10.1007/JHEP06(2012)085} {\bibfield  {journal} {\bibinfo  {journal}
  {JHEP}\ }\textbf {\bibinfo {volume} {1206}},\ \bibinfo {pages} {085}
  (\bibinfo {year} {2012}{\natexlab{b}})},\ \Eprint
  {http://arxiv.org/abs/1202.1986} {arXiv:1202.1986 [hep-th]} \BibitemShut
  {NoStop}%
%%CITATION = ARXIV:1202.1986;%%
\bibitem [{\citenamefont {Khosravi}\ \emph
  {et~al.}(2012{\natexlab{b}})\citenamefont {Khosravi}, \citenamefont
  {Sepangi},\ and\ \citenamefont {Shahidi}}]{Khosravi:2012rk}%
  \BibitemOpen
  \bibfield  {author} {\bibinfo {author} {\bibfnamefont {N.}~\bibnamefont
  {Khosravi}}, \bibinfo {author} {\bibfnamefont {H.~R.}\ \bibnamefont
  {Sepangi}}, \ and\ \bibinfo {author} {\bibfnamefont {S.}~\bibnamefont
  {Shahidi}},\ }\href {\doibase 10.1103/PhysRevD.86.043517} {\bibfield
  {journal} {\bibinfo  {journal} {Phys.Rev.}\ }\textbf {\bibinfo {volume}
  {D86}},\ \bibinfo {pages} {043517} (\bibinfo {year} {2012}{\natexlab{b}})},\
  \Eprint {http://arxiv.org/abs/1202.2767} {arXiv:1202.2767 [gr-qc]}
  \BibitemShut {NoStop}%
%%CITATION = ARXIV:1202.2767;%%
\bibitem [{\citenamefont {Berg}\ \emph {et~al.}(2012)\citenamefont {Berg},
  \citenamefont {Buchberger}, \citenamefont {Enander}, \citenamefont
  {M{\"o}rtsell},\ and\ \citenamefont {Sj{\"o}rs}}]{Berg:2012kn}%
  \BibitemOpen
  \bibfield  {author} {\bibinfo {author} {\bibfnamefont {M.}~\bibnamefont
  {Berg}}, \bibinfo {author} {\bibfnamefont {I.}~\bibnamefont {Buchberger}},
  \bibinfo {author} {\bibfnamefont {J.}~\bibnamefont {Enander}}, \bibinfo
  {author} {\bibfnamefont {E.}~\bibnamefont {M{\"o}rtsell}}, \ and\ \bibinfo
  {author} {\bibfnamefont {S.}~\bibnamefont {Sj{\"o}rs}},\ }\href {\doibase
  10.1088/1475-7516/2012/12/021} {\bibfield  {journal} {\bibinfo  {journal}
  {JCAP}\ }\textbf {\bibinfo {volume} {1212}},\ \bibinfo {pages} {021}
  (\bibinfo {year} {2012})},\ \Eprint {http://arxiv.org/abs/1206.3496}
  {arXiv:1206.3496 [gr-qc]} \BibitemShut {NoStop}%
%%CITATION = ARXIV:1206.3496;%%
\bibitem [{\citenamefont {Akrami}\ \emph
  {et~al.}(2013{\natexlab{b}})\citenamefont {Akrami}, \citenamefont
  {Koivisto},\ and\ \citenamefont {Sandstad}}]{Akrami:2012vf}%
  \BibitemOpen
  \bibfield  {author} {\bibinfo {author} {\bibfnamefont {Y.}~\bibnamefont
  {Akrami}}, \bibinfo {author} {\bibfnamefont {T.~S.}\ \bibnamefont
  {Koivisto}}, \ and\ \bibinfo {author} {\bibfnamefont {M.}~\bibnamefont
  {Sandstad}},\ }\href {\doibase 10.1007/JHEP03(2013)099} {\bibfield  {journal}
  {\bibinfo  {journal} {JHEP}\ }\textbf {\bibinfo {volume} {1303}},\ \bibinfo
  {pages} {099} (\bibinfo {year} {2013}{\natexlab{b}})},\ \Eprint
  {http://arxiv.org/abs/1209.0457} {arXiv:1209.0457 [astro-ph.CO]} \BibitemShut
  {NoStop}%
%%CITATION = ARXIV:1209.0457;%%
\bibitem [{\citenamefont {Akrami}\ \emph
  {et~al.}(2013{\natexlab{c}})\citenamefont {Akrami}, \citenamefont
  {Koivisto},\ and\ \citenamefont {Sandstad}}]{Akrami:2013pna}%
  \BibitemOpen
  \bibfield  {author} {\bibinfo {author} {\bibfnamefont {Y.}~\bibnamefont
  {Akrami}}, \bibinfo {author} {\bibfnamefont {T.~S.}\ \bibnamefont
  {Koivisto}}, \ and\ \bibinfo {author} {\bibfnamefont {M.}~\bibnamefont
  {Sandstad}},\ }\href@noop {} {\  (\bibinfo {year} {2013}{\natexlab{c}})},\
  \Eprint {http://arxiv.org/abs/1302.5268} {arXiv:1302.5268 [astro-ph.CO]}
  \BibitemShut {NoStop}%
%%CITATION = ARXIV:1302.5268;%%
\bibitem [{\citenamefont {Enander}\ and\ \citenamefont
  {M{\"o}rtsell}(2013)}]{Enander:2013kza}%
  \BibitemOpen
  \bibfield  {author} {\bibinfo {author} {\bibfnamefont {J.}~\bibnamefont
  {Enander}}\ and\ \bibinfo {author} {\bibfnamefont {E.}~\bibnamefont
  {M{\"o}rtsell}},\ }\href {\doibase 10.1007/JHEP10(2013)031} {\bibfield
  {journal} {\bibinfo  {journal} {JHEP}\ }\textbf {\bibinfo {volume} {1310}},\
  \bibinfo {pages} {031} (\bibinfo {year} {2013})},\ \Eprint
  {http://arxiv.org/abs/1306.1086} {arXiv:1306.1086 [astro-ph.CO]} \BibitemShut
  {NoStop}%
%%CITATION = ARXIV:1306.1086;%%
\bibitem [{\citenamefont {Fasiello}\ and\ \citenamefont
  {Tolley}(2013)}]{Fasiello:2013woa}%
  \BibitemOpen
  \bibfield  {author} {\bibinfo {author} {\bibfnamefont {M.}~\bibnamefont
  {Fasiello}}\ and\ \bibinfo {author} {\bibfnamefont {A.~J.}\ \bibnamefont
  {Tolley}},\ }\href {\doibase 10.1088/1475-7516/2013/12/002} {\bibfield
  {journal} {\bibinfo  {journal} {JCAP}\ }\textbf {\bibinfo {volume} {1312}},\
  \bibinfo {pages} {002} (\bibinfo {year} {2013})},\ \Eprint
  {http://arxiv.org/abs/1308.1647} {arXiv:1308.1647 [hep-th]} \BibitemShut
  {NoStop}%
%%CITATION = ARXIV:1308.1647;%%
\bibitem [{\citenamefont {K{\"o}nnig}\ \emph
  {et~al.}(2014{\natexlab{a}})\citenamefont {K{\"o}nnig}, \citenamefont
  {Patil},\ and\ \citenamefont {Amendola}}]{Konnig:2013gxa}%
  \BibitemOpen
  \bibfield  {author} {\bibinfo {author} {\bibfnamefont {F.}~\bibnamefont
  {K{\"o}nnig}}, \bibinfo {author} {\bibfnamefont {A.}~\bibnamefont {Patil}}, \
  and\ \bibinfo {author} {\bibfnamefont {L.}~\bibnamefont {Amendola}},\ }\href
  {\doibase 10.1088/1475-7516/2014/03/029} {\bibfield  {journal} {\bibinfo
  {journal} {JCAP}\ }\textbf {\bibinfo {volume} {1403}},\ \bibinfo {pages}
  {029} (\bibinfo {year} {2014}{\natexlab{a}})},\ \Eprint
  {http://arxiv.org/abs/1312.3208} {arXiv:1312.3208 [astro-ph.CO]} \BibitemShut
  {NoStop}%
%%CITATION = ARXIV:1312.3208;%%
\bibitem [{\citenamefont {K{\"o}nnig}\ and\ \citenamefont
  {Amendola}(2014)}]{Konnig:2014dna}%
  \BibitemOpen
  \bibfield  {author} {\bibinfo {author} {\bibfnamefont {F.}~\bibnamefont
  {K{\"o}nnig}}\ and\ \bibinfo {author} {\bibfnamefont {L.}~\bibnamefont
  {Amendola}},\ }\href {\doibase 10.1103/PhysRevD.90.044030} {\bibfield
  {journal} {\bibinfo  {journal} {Phys.Rev.}\ }\textbf {\bibinfo {volume}
  {D90}},\ \bibinfo {pages} {044030} (\bibinfo {year} {2014})},\ \Eprint
  {http://arxiv.org/abs/1402.1988} {arXiv:1402.1988 [astro-ph.CO]} \BibitemShut
  {NoStop}%
%%CITATION = ARXIV:1402.1988;%%
\bibitem [{\citenamefont {Comelli}\ \emph
  {et~al.}(2014{\natexlab{b}})\citenamefont {Comelli}, \citenamefont
  {Crisostomi},\ and\ \citenamefont {Pilo}}]{Comelli:2014bqa}%
  \BibitemOpen
  \bibfield  {author} {\bibinfo {author} {\bibfnamefont {D.}~\bibnamefont
  {Comelli}}, \bibinfo {author} {\bibfnamefont {M.}~\bibnamefont {Crisostomi}},
  \ and\ \bibinfo {author} {\bibfnamefont {L.}~\bibnamefont {Pilo}},\
  }\href@noop {} {\  (\bibinfo {year} {2014}{\natexlab{b}})},\ \Eprint
  {http://arxiv.org/abs/1403.5679} {arXiv:1403.5679 [hep-th]} \BibitemShut
  {NoStop}%
%%CITATION = ARXIV:1403.5679;%%
\bibitem [{\citenamefont {De~Felice}\ \emph {et~al.}(2014)\citenamefont
  {De~Felice}, \citenamefont {G{\"u}mr{\"u}k{\c c}{\"u}o{\u g}lu},
  \citenamefont {Mukohyama}, \citenamefont {Tanahashi},\ and\ \citenamefont
  {Tanaka}}]{DeFelice:2014nja}%
  \BibitemOpen
  \bibfield  {author} {\bibinfo {author} {\bibfnamefont {A.}~\bibnamefont
  {De~Felice}}, \bibinfo {author} {\bibfnamefont {A.~E.}\ \bibnamefont
  {G{\"u}mr{\"u}k{\c c}{\"u}o{\u g}lu}}, \bibinfo {author} {\bibfnamefont
  {S.}~\bibnamefont {Mukohyama}}, \bibinfo {author} {\bibfnamefont
  {N.}~\bibnamefont {Tanahashi}}, \ and\ \bibinfo {author} {\bibfnamefont
  {T.}~\bibnamefont {Tanaka}},\ }\href {\doibase 10.1088/1475-7516/2014/06/037}
  {\bibfield  {journal} {\bibinfo  {journal} {JCAP}\ }\textbf {\bibinfo
  {volume} {1406}},\ \bibinfo {pages} {037} (\bibinfo {year} {2014})},\ \Eprint
  {http://arxiv.org/abs/1404.0008} {arXiv:1404.0008 [hep-th]} \BibitemShut
  {NoStop}%
%%CITATION = ARXIV:1404.0008;%%
\bibitem [{\citenamefont {Solomon}\ \emph
  {et~al.}(2014{\natexlab{b}})\citenamefont {Solomon}, \citenamefont {Akrami},\
  and\ \citenamefont {Koivisto}}]{Solomon:2014dua}%
  \BibitemOpen
  \bibfield  {author} {\bibinfo {author} {\bibfnamefont {A.~R.}\ \bibnamefont
  {Solomon}}, \bibinfo {author} {\bibfnamefont {Y.}~\bibnamefont {Akrami}}, \
  and\ \bibinfo {author} {\bibfnamefont {T.~S.}\ \bibnamefont {Koivisto}},\
  }\href {\doibase 10.1088/1475-7516/2014/10/066} {\bibfield  {journal}
  {\bibinfo  {journal} {JCAP}\ }\textbf {\bibinfo {volume} {1410}},\ \bibinfo
  {pages} {066} (\bibinfo {year} {2014}{\natexlab{b}})},\ \Eprint
  {http://arxiv.org/abs/1404.4061} {arXiv:1404.4061 [astro-ph.CO]} \BibitemShut
  {NoStop}%
%%CITATION = ARXIV:1404.4061;%%
\bibitem [{\citenamefont {K{\"o}nnig}\ \emph
  {et~al.}(2014{\natexlab{b}})\citenamefont {K{\"o}nnig}, \citenamefont
  {Akrami}, \citenamefont {Amendola}, \citenamefont {Motta},\ and\
  \citenamefont {Solomon}}]{Konnig:2014xva}%
  \BibitemOpen
  \bibfield  {author} {\bibinfo {author} {\bibfnamefont {F.}~\bibnamefont
  {K{\"o}nnig}}, \bibinfo {author} {\bibfnamefont {Y.}~\bibnamefont {Akrami}},
  \bibinfo {author} {\bibfnamefont {L.}~\bibnamefont {Amendola}}, \bibinfo
  {author} {\bibfnamefont {M.}~\bibnamefont {Motta}}, \ and\ \bibinfo {author}
  {\bibfnamefont {A.~R.}\ \bibnamefont {Solomon}},\ }\href {\doibase
  10.1103/PhysRevD.90.124014} {\bibfield  {journal} {\bibinfo  {journal}
  {Phys.Rev.}\ }\textbf {\bibinfo {volume} {D90}},\ \bibinfo {pages} {124014}
  (\bibinfo {year} {2014}{\natexlab{b}})},\ \Eprint
  {http://arxiv.org/abs/1407.4331} {arXiv:1407.4331 [astro-ph.CO]} \BibitemShut
  {NoStop}%
%%CITATION = ARXIV:1407.4331;%%
\bibitem [{\citenamefont {Lagos}\ and\ \citenamefont
  {Ferreira}(2014)}]{Lagos:2014lca}%
  \BibitemOpen
  \bibfield  {author} {\bibinfo {author} {\bibfnamefont {M.}~\bibnamefont
  {Lagos}}\ and\ \bibinfo {author} {\bibfnamefont {P.~G.}\ \bibnamefont
  {Ferreira}},\ }\href@noop {} {\  (\bibinfo {year} {2014})},\ \Eprint
  {http://arxiv.org/abs/1410.0207} {arXiv:1410.0207 [gr-qc]} \BibitemShut
  {NoStop}%
%%CITATION = ARXIV:1410.0207;%%
\bibitem [{\citenamefont {Cusin}\ \emph {et~al.}(2014)\citenamefont {Cusin},
  \citenamefont {Durrer}, \citenamefont {Guarato},\ and\ \citenamefont
  {Motta}}]{Cusin:2014psa}%
  \BibitemOpen
  \bibfield  {author} {\bibinfo {author} {\bibfnamefont {G.}~\bibnamefont
  {Cusin}}, \bibinfo {author} {\bibfnamefont {R.}~\bibnamefont {Durrer}},
  \bibinfo {author} {\bibfnamefont {P.}~\bibnamefont {Guarato}}, \ and\
  \bibinfo {author} {\bibfnamefont {M.}~\bibnamefont {Motta}},\ }\href@noop {}
  {\  (\bibinfo {year} {2014})},\ \Eprint {http://arxiv.org/abs/1412.5979}
  {arXiv:1412.5979 [astro-ph.CO]} \BibitemShut {NoStop}%
%%CITATION = ARXIV:1412.5979;%%
\bibitem [{\citenamefont {Enander}\ \emph {et~al.}(2015)\citenamefont
  {Enander}, \citenamefont {Akrami}, \citenamefont {Mortsell}, \citenamefont
  {Renneby},\ and\ \citenamefont {Solomon}}]{Enander:2015vja}%
  \BibitemOpen
  \bibfield  {author} {\bibinfo {author} {\bibfnamefont {J.}~\bibnamefont
  {Enander}}, \bibinfo {author} {\bibfnamefont {Y.}~\bibnamefont {Akrami}},
  \bibinfo {author} {\bibfnamefont {E.}~\bibnamefont {Mortsell}}, \bibinfo
  {author} {\bibfnamefont {M.}~\bibnamefont {Renneby}}, \ and\ \bibinfo
  {author} {\bibfnamefont {A.~R.}\ \bibnamefont {Solomon}},\ }\href@noop {} {\
  (\bibinfo {year} {2015})},\ \Eprint {http://arxiv.org/abs/1501.02140}
  {arXiv:1501.02140 [astro-ph.CO]} \BibitemShut {NoStop}%
%%CITATION = ARXIV:1501.02140;%%
\bibitem [{\citenamefont {Enander}\ \emph {et~al.}(2014)\citenamefont
  {Enander}, \citenamefont {Solomon}, \citenamefont {Akrami},\ and\
  \citenamefont {M{\"o}rtsell}}]{Enander:2014xga}%
  \BibitemOpen
  \bibfield  {author} {\bibinfo {author} {\bibfnamefont {J.}~\bibnamefont
  {Enander}}, \bibinfo {author} {\bibfnamefont {A.~R.}\ \bibnamefont
  {Solomon}}, \bibinfo {author} {\bibfnamefont {Y.}~\bibnamefont {Akrami}}, \
  and\ \bibinfo {author} {\bibfnamefont {E.}~\bibnamefont {M{\"o}rtsell}},\
  }\href@noop {} {\  (\bibinfo {year} {2014})},\ \Eprint
  {http://arxiv.org/abs/1409.2860} {arXiv:1409.2860 [astro-ph.CO]} \BibitemShut
  {NoStop}%
%%CITATION = ARXIV:1409.2860;%%
\bibitem [{\citenamefont {Schmidt-May}(2014)}]{Schmidt-May:2014xla}%
  \BibitemOpen
  \bibfield  {author} {\bibinfo {author} {\bibfnamefont {A.}~\bibnamefont
  {Schmidt-May}},\ }\href@noop {} {\  (\bibinfo {year} {2014})},\ \Eprint
  {http://arxiv.org/abs/1409.3146} {arXiv:1409.3146 [gr-qc]} \BibitemShut
  {NoStop}%
%%CITATION = ARXIV:1409.3146;%%
\bibitem [{\citenamefont {Comelli}\ \emph {et~al.}(2015)\citenamefont
  {Comelli}, \citenamefont {Crisostomi}, \citenamefont {Koyama}, \citenamefont
  {Pilo},\ and\ \citenamefont {Tasinato}}]{Comelli:2015pua}%
  \BibitemOpen
  \bibfield  {author} {\bibinfo {author} {\bibfnamefont {D.}~\bibnamefont
  {Comelli}}, \bibinfo {author} {\bibfnamefont {M.}~\bibnamefont {Crisostomi}},
  \bibinfo {author} {\bibfnamefont {K.}~\bibnamefont {Koyama}}, \bibinfo
  {author} {\bibfnamefont {L.}~\bibnamefont {Pilo}}, \ and\ \bibinfo {author}
  {\bibfnamefont {G.}~\bibnamefont {Tasinato}},\ }\href@noop {} {\  (\bibinfo
  {year} {2015})},\ \Eprint {http://arxiv.org/abs/1501.00864} {arXiv:1501.00864
  [hep-th]} \BibitemShut {NoStop}%
%%CITATION = ARXIV:1501.00864;%%
\bibitem [{\citenamefont {Gumrukcuoglu}\ \emph {et~al.}(2015)\citenamefont
  {Gumrukcuoglu}, \citenamefont {Heisenberg}, \citenamefont {Mukohyama},\ and\
  \citenamefont {Tanahashi}}]{Gumrukcuoglu:2015nua}%
  \BibitemOpen
  \bibfield  {author} {\bibinfo {author} {\bibfnamefont {A.~E.}\ \bibnamefont
  {Gumrukcuoglu}}, \bibinfo {author} {\bibfnamefont {L.}~\bibnamefont
  {Heisenberg}}, \bibinfo {author} {\bibfnamefont {S.}~\bibnamefont
  {Mukohyama}}, \ and\ \bibinfo {author} {\bibfnamefont {N.}~\bibnamefont
  {Tanahashi}},\ }\href@noop {} {\  (\bibinfo {year} {2015})},\ \Eprint
  {http://arxiv.org/abs/1501.02790} {arXiv:1501.02790 [hep-th]} \BibitemShut
  {NoStop}%
%%CITATION = ARXIV:1501.02790;%%
\bibitem [{\citenamefont {K{\"o}nnig}(2015)}]{Konnig:2015lfa}%
  \BibitemOpen
  \bibfield  {author} {\bibinfo {author} {\bibfnamefont {F.}~\bibnamefont
  {K{\"o}nnig}},\ }\href@noop {} {\  (\bibinfo {year} {2015})},\ \Eprint
  {http://arxiv.org/abs/1503.07436} {arXiv:1503.07436 [astro-ph.CO]}
  \BibitemShut {NoStop}%
%%CITATION = ARXIV:1503.07436;%%
\bibitem [{\citenamefont {Amendola}\ \emph {et~al.}(2015)\citenamefont
  {Amendola}, \citenamefont {Koennig}, \citenamefont {Martinelli},
  \citenamefont {Pettorino},\ and\ \citenamefont
  {Zumalacarregui}}]{Amendola:2015tua}%
  \BibitemOpen
  \bibfield  {author} {\bibinfo {author} {\bibfnamefont {L.}~\bibnamefont
  {Amendola}}, \bibinfo {author} {\bibfnamefont {F.}~\bibnamefont {Koennig}},
  \bibinfo {author} {\bibfnamefont {M.}~\bibnamefont {Martinelli}}, \bibinfo
  {author} {\bibfnamefont {V.}~\bibnamefont {Pettorino}}, \ and\ \bibinfo
  {author} {\bibfnamefont {M.}~\bibnamefont {Zumalacarregui}},\ }\href@noop {}
  {\  (\bibinfo {year} {2015})},\ \Eprint {http://arxiv.org/abs/1503.02490}
  {arXiv:1503.02490 [astro-ph.CO]} \BibitemShut {NoStop}%
%%CITATION = ARXIV:1503.02490;%%
\bibitem [{\citenamefont {Johnson}\ and\ \citenamefont
  {Terrana}(2015)}]{Johnson:2015tfa}%
  \BibitemOpen
  \bibfield  {author} {\bibinfo {author} {\bibfnamefont {M.}~\bibnamefont
  {Johnson}}\ and\ \bibinfo {author} {\bibfnamefont {A.}~\bibnamefont
  {Terrana}},\ }\href@noop {} {\  (\bibinfo {year} {2015})},\ \Eprint
  {http://arxiv.org/abs/1503.05560} {arXiv:1503.05560 [astro-ph.CO]}
  \BibitemShut {NoStop}%
%%CITATION = ARXIV:1503.05560;%%
\bibitem [{\citenamefont {Maeda}\ and\ \citenamefont
  {Volkov}(2013)}]{Maeda:2013bha}%
  \BibitemOpen
  \bibfield  {author} {\bibinfo {author} {\bibfnamefont {K.-i.}\ \bibnamefont
  {Maeda}}\ and\ \bibinfo {author} {\bibfnamefont {M.~S.}\ \bibnamefont
  {Volkov}},\ }\href {\doibase 10.1103/PhysRevD.87.104009} {\bibfield
  {journal} {\bibinfo  {journal} {Phys.Rev.}\ }\textbf {\bibinfo {volume}
  {D87}},\ \bibinfo {pages} {104009} (\bibinfo {year} {2013})},\ \Eprint
  {http://arxiv.org/abs/1302.6198} {arXiv:1302.6198 [hep-th]} \BibitemShut
  {NoStop}%
%%CITATION = ARXIV:1302.6198;%%
\bibitem [{\citenamefont {Akrami}\ \emph {et~al.}(2015)\citenamefont {Akrami},
  \citenamefont {Hassan}, \citenamefont {K{\"o}nnig}, \citenamefont
  {Schmidt-May},\ and\ \citenamefont {Solomon}}]{Akrami:2015qga}%
  \BibitemOpen
  \bibfield  {author} {\bibinfo {author} {\bibfnamefont {Y.}~\bibnamefont
  {Akrami}}, \bibinfo {author} {\bibfnamefont {S.}~\bibnamefont {Hassan}},
  \bibinfo {author} {\bibfnamefont {F.}~\bibnamefont {K{\"o}nnig}}, \bibinfo
  {author} {\bibfnamefont {A.}~\bibnamefont {Schmidt-May}}, \ and\ \bibinfo
  {author} {\bibfnamefont {A.~R.}\ \bibnamefont {Solomon}},\ }\href@noop {} {\
  (\bibinfo {year} {2015})},\ \Eprint {http://arxiv.org/abs/1503.07521}
  {arXiv:1503.07521 [gr-qc]} \BibitemShut {NoStop}%
%%CITATION = ARXIV:1503.07521;%%
\bibitem [{\citenamefont {Bolejko}\ and\ \citenamefont
  {Lasky}(2008)}]{Bolejko:2008ya}%
  \BibitemOpen
  \bibfield  {author} {\bibinfo {author} {\bibfnamefont {K.}~\bibnamefont
  {Bolejko}}\ and\ \bibinfo {author} {\bibfnamefont {P.}~\bibnamefont
  {Lasky}},\ }\href {\doibase 10.1111/j.1745-3933.2008.00555.x} {\bibfield
  {journal} {\bibinfo  {journal} {Mon.Not.Roy.Astron.Soc.}\ }\textbf {\bibinfo
  {volume} {391}},\ \bibinfo {pages} {59} (\bibinfo {year} {2008})},\ \Eprint
  {http://arxiv.org/abs/0809.0334} {arXiv:0809.0334 [astro-ph]} \BibitemShut
  {NoStop}%
%%CITATION = ARXIV:0809.0334;%%
\bibitem [{\citenamefont {Bolejko}\ \emph {et~al.}(2011)\citenamefont
  {Bolejko}, \citenamefont {Celerier},\ and\ \citenamefont
  {Krasinski}}]{Bolejko:2011jc}%
  \BibitemOpen
  \bibfield  {author} {\bibinfo {author} {\bibfnamefont {K.}~\bibnamefont
  {Bolejko}}, \bibinfo {author} {\bibfnamefont {M.-N.}\ \bibnamefont
  {Celerier}}, \ and\ \bibinfo {author} {\bibfnamefont {A.}~\bibnamefont
  {Krasinski}},\ }\href {\doibase 10.1088/0264-9381/28/16/164002} {\bibfield
  {journal} {\bibinfo  {journal} {Class.Quant.Grav.}\ }\textbf {\bibinfo
  {volume} {28}},\ \bibinfo {pages} {164002} (\bibinfo {year} {2011})},\
  \Eprint {http://arxiv.org/abs/1102.1449} {arXiv:1102.1449 [astro-ph.CO]}
  \BibitemShut {NoStop}%
%%CITATION = ARXIV:1102.1449;%%
\bibitem [{\citenamefont {Lemaitre}(1997)}]{Lemaitre:1933gd}%
  \BibitemOpen
  \bibfield  {author} {\bibinfo {author} {\bibfnamefont {G.}~\bibnamefont
  {Lemaitre}},\ }\href {\doibase 10.1023/A:1018855621348} {\bibfield  {journal}
  {\bibinfo  {journal} {Gen.Rel.Grav.}\ }\textbf {\bibinfo {volume} {29}},\
  \bibinfo {pages} {641} (\bibinfo {year} {1997})}\BibitemShut {NoStop}%
%%CITATION = GRGVA,29,641;%%
\bibitem [{\citenamefont {Lasky}\ and\ \citenamefont
  {Bolejko}(2010)}]{Lasky:2010vn}%
  \BibitemOpen
  \bibfield  {author} {\bibinfo {author} {\bibfnamefont {P.~D.}\ \bibnamefont
  {Lasky}}\ and\ \bibinfo {author} {\bibfnamefont {K.}~\bibnamefont
  {Bolejko}},\ }\href {\doibase 10.1088/0264-9381/27/3/035011} {\bibfield
  {journal} {\bibinfo  {journal} {Class.Quant.Grav.}\ }\textbf {\bibinfo
  {volume} {27}},\ \bibinfo {pages} {035011} (\bibinfo {year} {2010})},\
  \Eprint {http://arxiv.org/abs/1001.1159} {arXiv:1001.1159 [astro-ph.CO]}
  \BibitemShut {NoStop}%
%%CITATION = ARXIV:1001.1159;%%
\bibitem [{\citenamefont {Harko}\ \emph {et~al.}(2014)\citenamefont {Harko},
  \citenamefont {Lobo},\ and\ \citenamefont {Mak}}]{Harko:2014nya}%
  \BibitemOpen
  \bibfield  {author} {\bibinfo {author} {\bibfnamefont {T.}~\bibnamefont
  {Harko}}, \bibinfo {author} {\bibfnamefont {F.~S.~N.}\ \bibnamefont {Lobo}},
  \ and\ \bibinfo {author} {\bibfnamefont {M.}~\bibnamefont {Mak}},\ }\href
  {\doibase 10.3390/galaxies2040496} {\bibfield  {journal} {\bibinfo  {journal}
  {Galaxies}\ }\textbf {\bibinfo {volume} {2}},\ \bibinfo {pages} {496}
  (\bibinfo {year} {2014})},\ \Eprint {http://arxiv.org/abs/1410.5213}
  {arXiv:1410.5213 [gr-qc]} \BibitemShut {NoStop}%
%%CITATION = ARXIV:1410.5213;%%
\bibitem [{\citenamefont {Saha}(2005)}]{Saha:2004mr}%
  \BibitemOpen
  \bibfield  {author} {\bibinfo {author} {\bibfnamefont {B.}~\bibnamefont
  {Saha}},\ }\href@noop {} {\bibfield  {journal} {\bibinfo  {journal}
  {Chin.J.Phys.}\ }\textbf {\bibinfo {volume} {43}},\ \bibinfo {pages} {1035}
  (\bibinfo {year} {2005})},\ \Eprint {http://arxiv.org/abs/gr-qc/0412078}
  {arXiv:gr-qc/0412078 [gr-qc]} \BibitemShut {NoStop}%
%%CITATION = GR-QC/0412078;%%
\bibitem [{\citenamefont {Mukhanov}\ \emph {et~al.}(1992)\citenamefont
  {Mukhanov}, \citenamefont {Feldman},\ and\ \citenamefont
  {Brandenberger}}]{Mukhanov:1990me}%
  \BibitemOpen
  \bibfield  {author} {\bibinfo {author} {\bibfnamefont {V.~F.}\ \bibnamefont
  {Mukhanov}}, \bibinfo {author} {\bibfnamefont {H.}~\bibnamefont {Feldman}}, \
  and\ \bibinfo {author} {\bibfnamefont {R.~H.}\ \bibnamefont
  {Brandenberger}},\ }\href@noop {} {\bibfield  {journal} {\bibinfo  {journal}
  {Phys.Rept.}\ } (\bibinfo {year} {1992})}\BibitemShut {NoStop}%
\bibitem [{\citenamefont {Lagos}\ \emph {et~al.}(2014)\citenamefont {Lagos},
  \citenamefont {Bañados}, \citenamefont {Ferreira},\ and\ \citenamefont
  {García-Sáenz}}]{Lagos:2013aua}%
  \BibitemOpen
  \bibfield  {author} {\bibinfo {author} {\bibfnamefont {M.}~\bibnamefont
  {Lagos}}, \bibinfo {author} {\bibfnamefont {M.}~\bibnamefont {Bañados}},
  \bibinfo {author} {\bibfnamefont {P.~G.}\ \bibnamefont {Ferreira}}, \ and\
  \bibinfo {author} {\bibfnamefont {S.}~\bibnamefont {García-Sáenz}},\ }\href
  {\doibase 10.1103/PhysRevD.89.024034} {\bibfield  {journal} {\bibinfo
  {journal} {Phys.Rev.}\ }\textbf {\bibinfo {volume} {D89}},\ \bibinfo {pages}
  {024034} (\bibinfo {year} {2014})},\ \Eprint {http://arxiv.org/abs/1311.3828}
  {arXiv:1311.3828 [gr-qc]} \BibitemShut {NoStop}%
%%CITATION = ARXIV:1311.3828;%%
\end{thebibliography}%

\end{document}